\documentclass[aps,prl,twocolumn,showpacs,preprintnumbers]{revtex4}
\usepackage{amsmath}
\usepackage{dcolumn}
\usepackage{bm}
\usepackage{subfigure}
\usepackage{amsfonts}
\usepackage{amssymb}
\usepackage{makeidx}
\usepackage{epsfig}
\usepackage{graphicx}

\begin{document}

\title{\textbf{Covariant formulation of spatially non-symmetric kinetic
equilibria\\
in magnetized astrophysical plasmas}}
\author{Claudio Cremaschini\thanks{%
Electronic-mail: claudiocremaschini@gmail.com}$^{a}$, Massimo Tessarotto$%
^{a,b} $ and Zden\v{e}k Stuchl\'{\i}k$^{a}$}
\affiliation{$^{a}$Institute of Physics, Faculty of Philosophy and Science, Silesian
University in Opava, Bezru\v{c}ovo n\'{a}m.13, CZ-74601 Opava, Czech Republic%
\\
$^{b}$Department of Mathematics and Geosciences, University of Trieste, Via
Valerio 12, 34127 Trieste, Italy}
\date{\today }

\begin{abstract}
Astrophysical plasmas in the surrounding of compact objects and subject to
intense gravitational and electromagnetic fields are believed to give rise
to relativistic regimes. Theoretical and observational evidence suggest that
magnetized plasmas of this type are collisionless and can persist for long
times (e.g., with respect to a distant observer, coordinate, time), while
exhibiting geometrical structures characterized by the absence of
well-defined spatial symmetries. In this paper the problem is posed whether
such configurations can correspond to some kind of kinetic equilibrium. The
issue is addressed from a theoretical perspective in the framework of a
covariant Vlasov statistical description, which relies on the method of
invariants. For this purpose, a systematic covariant variational formulation
of gyrokinetic theory is developed, which holds without requiring any
symmetry condition on the background fields. As a result, an asymptotic
representation of the relativistic particle magnetic moment is obtained from
its formal exact solution, in terms of a suitably-defined invariant series
expansion parameter (perturbative representation). On such a basis it is
shown that spatially non-symmetric kinetic equilibria can actually be
determined, an example being provided by Gaussian-like distributions. As an
application, the physical mechanisms related to the occurrence of a
non-vanishing equilibrium fluid 4-flow are investigated.
\end{abstract}

\pacs{05.20.Dd, 52.25.Xz, 52.27.Ny, 52.30.Gz, }
\maketitle

\bigskip


\section{Introduction}

A fundamental theoretical problem in astrophysics concerns the description
of the complex phenomenology of plasmas arising in the surrounding of
compact objects, such as in accretion-disc scenarios around black holes or
neutron stars and in relativistic jets related to them. For these systems,
both single-particle and macroscopic fluid velocities of the plasma can
become relativistic, i.e. comparable with the speed of light $c$ when
measured in a defined reference frame (e.g., the laboratory or co-moving
frames), while also space-time curvature effects associated with strong
gravitational fields can be relevant in the plasma configuration domain.
When these circumstances are relevant, relativistic covariant approaches
need to be adopted. For astrophysical plasmas, this requires the proper
formulation of the covariant description of charge and fluid dynamics
subject to the simultaneous action of electromagnetic (EM) and gravitational
fields, determined respectively by the Maxwell equations and by the Einstein
equations of General Relativity \cite{LL}.

As far as the plasma is concerned, different statistical descriptions can be
implemented in principle for the study of the corresponding $N-$body system
formed by species charged particles, with $N\gg 1$. These include kinetic
theories and fluid or magnetohydrodynamic (MHD) treatments, with the choice
depending both on the type of phenomena to be studied and on the properties
of the plasma itself \cite{Cr2010}. This in turn requires the identification
of the regimes that characterize the plasma states, which are determined by
the plasma microscopic and macroscopic properties \cite{Cr2012}. Two types
of processes should be distinguished in this regard: the first one concerns
particle dynamics, while the second\ one deals with collective phenomena in
plasmas. In particular, the first category includes a variety of EM
interactions/processes ranging from particle binary Coulomb collisions,
single-particle radiation emission (radiation-reaction) \cite%
{maha-1,maha-2,EPJ1,EPJ2,EPJ3,EPJ4,EPJ5,EPJ6}, chemical or nuclear processes
as well as the influence of intense electromagnetic fields on single
particle dynamics (gyrokinetic theory). In the second group it is worth
mentioning ubiquitous phenomena like plasma generation by feeding or
ionization processes, formation and stability of disc and jet plasmas \cite%
{cz1}, quasi-periodic oscillations, radiation phenomena associated with
external sources \cite%
{as7,as8,as2,as3,as1,as9,as10,as6,as4,ks1,ks2,ks3,ks4,ks5,ks6,ks7},
collective effects of magnetic trapping and\ plasma\textbf{\ }confinement,
plasma thermal properties (typically related to the prescription of the
fluid pressure tensor), local plasma flows such as those occurring in the
vicinity of compact objects, etc.

The prohibitive complexity of plasma phenomenology makes impossible the
development of a unified theory. This demands in fact the identification of
specific subsets of phenomena to be investigated. In the following we
consider strongly-magnetized collisionless plasmas which can be treated in
the framework of a covariant Vlasov-Maxwell formulation. In particular, in
this paper we focus our attention on the relativistic plasmas in which
single-particle dynamics can be also regarded as relativistic, as it can be
relevant in relativistic jets, central regions of accretion discs and more
generally, for plasmas occurring in the vicinity of massive objects. The
precise definition of the applicable plasma regimes will be discussed below.
However, we stress that the present theory actually relies only on the
prescription of appropriate asymptotic conditions holding for
single-particle dynamics, to be defined with respect to a suitable reference
frame.

Experimental evidence as well theoretical studies suggest that in the
astrophysical environments considered here multi-species magnetized and
collisionless plasmas arise, which are locally or globally relativistic in
the sense indicated above, at least in particular subsets of the
configuration domains (see for example Refs.\cite{co2,co5,co3,co1,co4}). For
systems of this kind, the appropriate statistical framework is provided by
the covariant Vlasov kinetic theory, to be generally supplemented by both
Maxwell and Einstein equations. This type of kinetic description allows for
both phase-space single-particle as well as (EM and gravitational)
collective system dynamics to be consistently taken into account \cite%
{degroot}. Within such a description, the fundamental quantity is
represented by the species kinetic distribution function (KDF) $f_{\mathrm{s}%
}$, where $s$ is the species index, whose dynamical evolution is determined
by the Vlasov equation (see details below).

A characteristic feature of the KDF is that it is defined in the
single-particle phase-space, so that it carries the information about the
plasma distribution in both velocity and configuration domains. As a
consequence, the prescription of the closure conditions for the fluid
equations becomes unique and self-consistent in this framework. In
particular, there is a number of cases where this happens in collisionless
plasmas for which kinetic effects must be properly accounted for. These
arise either due to single-particle dynamics (for example, gyrokinetic
dynamics, occurrence of adiabatic invariants, particle trapping, etc.), or
due to the peculiar form of the KDF, which may depart from an isotropic
local Maxwellian. In such a case, in fact, one expects the fluid closure
conditions to become non-trivial. This occurs in particular in the presence
of temperature and pressure anisotropies, local flows, such as diamagnetic
flow velocities, finite Larmor-radius (FLR) and energy-correction effects 
\cite{Cr2011,Cr2011a,PRL,Cr2013,Cr2013b,Cr2013c,PRE-new,Catto1987}.

A qualitative feature of astrophysical magnetized plasmas is related to the
occurrence of kinetic plasma regimes which persist for long times (with
respect to the observer and/or plasma characteristic times), despite the
presence of macroscopic time-varying phenomena of various origin, such as
flows, non-uniform gravitational/EM fields and EM radiation \cite{Volpi},
possibly including that arising from single-particle radiation-reaction
processes \cite{maha-1,maha-2,EPJ1,EPJ2,EPJ3,EPJ4,EPJ5,EPJ6}. It is argued
that, for collisionless plasmas, these states might actually correspond - at
least locally and in a suitable asymptotic sense - to some kind of kinetic
equilibrium which characterizes the species KDFs. This is realized when the
latter distributions are all assumed to be functions only of the
single-particle adiabatic invariants. Therefore, in this sense kinetic
equilibria may arise also in physical scenarios in which macroscopic fluid
fields (e.g., fluid 4-flows) and/or the EM field might be time dependent
when observed from an observer reference frame. It is worth noting that in
practice this may require only that the corresponding KDF actually changes
slowly in time, with respect to the microscopic Larmor time-scales, so that
from the phenomenological viewpoint these states can still be regarded as
quasi-stationary or even non-stationary. The practical consequences of this
may be far reaching since kinetic theory might actually afford to overcome
apparently unsolvable theoretical challenges in this connection.

For non-relativistic systems, the subject was investigated systematically in
a series of recent works (see Refs. \cite%
{Cr2010,Cr2012,Cr2011,Cr2011a,PRL,Cr2013,Cr2013b,Cr2013c,PRE-new,APJS}),
where equilibrium solutions were constructed in the framework of the
Vlasov-Maxwell description regarding non-relativistic kinetic equilibria and
their stability properties in the case of collisionless magnetized plasmas
subject to stationary or quasi-stationary EM and gravitational fields. To
this aim, a method\ based on the identification of the relevant plasma
regimes \cite{Cr2012} and on the use of particle invariants was implemented.
In particular, a major achievement of such an approach is the discovery of
axisymmetric kinetic equilibria which exhibit a number of peculiar physical
properties, possibly akin to experimental observation, which range from
quasi-neutrality, the self-generation of the equilibrium EM fields and the
production of macroscopic azimuthal and poloidal flow velocities, together
with anisotropic temperature and pressure. Remarkably, these equilibria,
which have been proved to exist in particular in the so-called \textit{%
strongly-magnetized plasma regime}, belong to a more general functional
class whose KDFs are all absolutely stable with respect to axisymmetric EM
perturbations, see Refs.\cite{PRL,PRE-new}.

Further interesting developments concern, however, a more general physical
setting in which some of the relevant symmetry properties characteristic of
the equilibria indicated above, may be in part lost. These include both
spatially non-symmetric kinetic equilibria in which energy is conserved \cite%
{Cr2013} as well as energy-independent kinetic equilibria \cite{PRE-new} in
which a continuous spatial symmetry of some kind still survives. Possible
physical realizations of kinetic equilibria of this type are in principle
ubiquitous. In fact, analogous conclusions apply in principle both to
astrophysical and laboratory systems \cite{Cr2011,Cr2011a}, ranging from
non-axisymmetric and rotating accretion-disc plasmas, localized
non-axisymmetric plasma flows in astrophysical plasmas (such as jets and
star flares), as well as Laboratory plasmas (such as Tokamak, RFP and
Stellarator systems), all characterized by the presence of a rich
phenomenology \cite{Cr2013c}.

The study of plasma dynamics in high-energy astrophysical scenarios has
gained interest in the last years among the scientific community, thanks to
the development of accurate and advanced numerical tools able to investigate
in greater details the physical phenomena occurring in these systems (see
for example Refs.\cite{pe-1,pe0,pe1,pe2,pe3,pe4,pe5,pe6}). In this regard,
recent general-relativistic theoretical investigations \cite%
{ma-1,ma-2,ma-3,ma-4} and MHD numerical simulations \cite{Re2,Re1,Re0,Re00}
have studied the properties of disc plasmas accreting onto compact objects
and interacting with them in combination with intense magnetic fields, and
the mechanisms that can finally lead to the generation of relativistic jets.
As a consequence of such a dynamical evolution, structures of this type show
the evidence of the development of non-axisymmetric morphologies \cite%
{na2,na3,na4,na5,na6,na1}. To be more precise, this refers to the behavior
of both the EM and fluid fields, while the background gravitational field
can still be regarded as being characterized by the occurrence of space-time
symmetries of some kind (e.g., to be defined with respect to the distant
observer coordinate system).\textbf{\ }Therefore, they correspond to
magnetized plasmas arising in strong gravitational fields and characterized
by absence of spatial symmetries.

In view of these considerations, in this paper the problem is posed whether
these configurations may be understood on the grounds of a kinetic
description, more precisely in terms of kinetic equilibria characterizing
relativistic collisionless plasmas in curved space-time. Besides
astrophysical applications, the problem has a more general interest since it
represents a challenge also in theoretical plasma physics. Indeed, an
interesting connection is represented by the already-mentioned recent
discovery of spatially non-symmetric kinetic equilibria \cite{Cr2013}. Such
a development solves a historical conjecture originally posed by Grad on the
very existence of toroidal non-symmetric equilibria \cite%
{Grad84,Grad67,Grad58}.

The possible generalization of these results to relativistic plasmas\ of the
type indicated above would represent a notable achievement for its
theoretical implications in high-energy astrophysics. To reach the goal,
extending the non-relativistic treatment developed in Ref.\cite{Cr2013}, the
problem must be couched in the\ context of\ covariant formulation of
gyrokinetic (GK) theory to hold also in curved space-time. In such a
context, the possibility of developing a systematic perturbative (or even
non-perturbative) approach, able to retain all physically-relevant
GK-effects, becomes essential (see definition below). In terms of a
Lagrangian variational description of single-particle dynamics, one expects
the accurate evaluation of the particle magnetic moment to become possible,
thus allowing one,\ in particular, to determine also higher-order
Larmor-radius corrections.

In the framework of a covariant formulation of GK theory, the reference
works are provided by Refs.\cite{Bek1,Bek,bek3}. In particular, starting
point of the present study is the novel implementation of a covariant
treatment of GK theory based on the explicit construction of an extended
phase-space transformation for the single-particle state. We stress that
such an extended transformation is in principle finite, so that it can
involve a non-perturbative formulation, a feature which allows one to
determine an exact representation of the magnetic moment \cite{bek3}.
Starting from this result, for practical applications, however, the adoption
of a perturbative theory becomes unavoidable. This consists in the
construction of an asymptotic representation for the relativistic particle
magnetic moment expressed in terms of a convergent series expansion, to be
denoted in the following as perturbative representation for the magnetic
moment. The latter is based on the introduction of a suitable 4-scalar
expansion parameter $\varepsilon $ related to the particle Larmor-radius
characteristic length whose mathematical definition is given for convenience
below in Eq.(\ref{epsi}). Although asymptotic in character (relative to the
Larmor-radius expansion), the perturbative theory is fundamental to display
higher-order velocity corrections (in terms of a Larmor-radius expansion)
carried by the magnetic moment. In particular, new contributions are
discovered in this way, which are missing in the customary leading-order
representation of the magnetic moment usually adopted in the literature \cite%
{Bek1,Bek}.

Although in principle the absence of spatial symmetries may complicate the
treatment of single-particle dynamics and its GK formulation, difficulties
can arise at the macroscopic level in connection with the statistical plasma
phenomenology. As discussed also in Ref.\cite{Cr2013} in the case of
non-relativistic equilibria, this concerns the possibility of recovering in
such a configuration the relevant physical properties which characterize
spatially-symmetric systems, including equilibrium mechanisms for the
occurrence of plasma flows, pressure and temperature anisotropies. In this
regard, the GK theory developed here is a necessary prerequisite for the
consistent formulation of the Vlasov kinetic theory. We stress that for
spatially non-symmetric equilibria the calculation of the magnetic moment to
higher accuracy beyond its leading-order expression, i.e. including at least
first-order terms in the expansion parameter $\varepsilon $, is demanded.
One of the main reasons for this requirement is related to the fact that, as
in the non-relativistic case, in the absence of global spatial symmetries,
the magnetic moment conservation, and in particular its higher-order
contributions, can provide a unique physical mechanism for the occurrence of
equilibrium flows in collisionless plasmas, both along the local direction
of magnetic field lines (parallel flow) and in the perpendicular directions.
Nevertheless, \textquotedblleft a priori\textquotedblright\ it is not
obvious that relativistic equilibria of this type should display qualitative
features analogous to those of non-relativistic systems. In particular, it
is important to ascertain what are the physical mechanisms responsible for
the generation of flows in relativistic kinetic equilibria. In addition, one
expects that also non-uniform fluid fields and temperature anisotropy may
appear in combination with non-vanishing fluid flows. Indeed, the
prerequisite for such a target is the correct determination of the
single-particle adiabatic invariants also in the relativistic context.

\section{Goals of the study}

Putting these issues in perspective, in this paper we carry out a covariant
generalization of the work presented in Ref.\cite{Cr2013}. The investigation
concerns first the formulation of both exact and perturbative single-charge
dynamics in the framework of a covariant GK theory and the derivation of the
corresponding expressions of the particle magnetic moment. The second goal
refers to the construction of spatially non-symmetric kinetic equilibria for
relativistic collisionless and magnetized plasmas in curved space-time, and
the identification of the kinetic mechanisms that are responsible for the
occurrence of non-vanishing local plasma flows.

The present work is mainly theoretically-oriented and focused on providing
some fundamentals concerning the theoretical description of relativistic
kinetic equilibria in collisionless plasmas based on a covariant formulation
of GK theory. No explicit applications to real astrophysical situations are
considered here, they will be subject of future separate treatments. In
particular, the goals of the study are detailed as follows:

1)\ To establish a non-perturbative covariant GK theory in terms of an
extended guiding-center transformation, i.e., involving both the particle
4-position and 4-velocity. The theory permits introduction of an exact
non-perturbative representation for the relativistic particle magnetic
moment $m^{\prime }$. A formalism based on the use of EM-tetrad reference
frame is adopted for this purpose. It is worth pointing out that the
non-perturbative approach developed here prevents the use of standard
perturbative methods, like the Lie-transform approach.\textbf{\ }For this
reason, a direct transformation method based on the adoption of
superabundant hybrid (i.e., non canonical) variables is implemented.

2)\ To develop an explicit perturbative treatment of covariant GK theory. In
particular, the perturbative Larmor-radius expansion is constructed by
imposing, as usual, that the fundamental Lagrangian differential form
becomes gyrophase-independent at each order \cite{Cr2013}. The perturbative
calculation is carried out through second order in a suitable 4-scalar
dimensionless expansion parameter $\varepsilon \ll 1$, namely retaining
contributions of $O\left( \varepsilon ^{0}\right) $\ and $O\left(
\varepsilon \right) $\ in the GK Lagrangian differential form. As a notable
feature, the asymptotic convergence of the theory is proved to hold for
strongly-magnetized plasmas in both strong and weak electric field regimes
defined below.

3) To determine an asymptotic expression for the relativistic 4-scalar
particle magnetic moment $m^{\prime }$ which includes corrections of $%
O\left( \varepsilon \right) $ according to the perturbative theory. In
analogy with the non-relativistic solution obtained in Ref.\cite{Cr2013},
the target here is to prove that the new perturbative expression obtained in
this way depends explicitly on the guiding-center particle 4-velocity
components. This feature in fact is essential for the kinetic treatment of
non-symmetric systems.

4) On the basis of this result, to prove that kinetic equilibrium solutions
of the Vlasov equation exist for relativistic plasmas in curved space-time.
Configurations corresponding to the validity of the stationarity condition
and absence of spatial symmetries (for the observer reference frame) are
treated. A family of admissible solutions is constructed by implementing the
method of invariants, which include in particular Gaussian-like KDFs.
Although such an approach is quite standard in the literature (see for
example Ref.\cite{ehl} for a general relativistic kinetic treatment of
neutral gases), its application to relativistic collisionless plasmas in the
framework of a covariant GK theory is revealing of the complex and
characteristic phenomenology which arises in these systems in combination
with non-uniform EM fields.

5)\ To proceed with the analytical estimate of the fluid 4-flow associated
with the kinetic equilibrium, to be evaluated in the EM-tetrad reference
frame \cite{Bek1,Bek}. Qualitative properties of the velocity integrals
involved in the definition of the 4-flow are determined, when the
equilibrium KDF is identified with the Gaussian-like KDF.

6)\ From the calculation of point 5), to prove on general grounds that these
equilibria generally give rise to non-vanishing local 4-flows. In
particular, the target here consists in determining the possible physical
mechanisms responsible for the generation of flows in non-symmetric
equilibria. These are shown to be related both to the velocity dependence
contained in the magnetic moment and to the conserved canonical momentum
which is conjugate to the observer time coordinate. In particular,\ the
existence of intrinsic relativistic effects is pointed out, which differ
from the corresponding non-relativistic limit treated in Ref.\cite{Cr2013}.

The scheme of the paper is the following one. In Section 3 we introduce the
EM-tetrad formalism. Sections 4 and 5 deal with the construction of a
non-perturbative and a perturbative covariant GK theory respectively. In
Section 6 the Lagrangian differential form in GK variables is obtained.
Sections 7 and 8 contain the leading-order and first-order perturbative
solutions of the GK theory, focusing in particular on deriving the
corresponding expressions for the particle magnetic moment. In Section 9 the
construction of relativistic kinetic equilibria for spatially non-symmetric
collisionless plasmas is addressed and an explicit realization of
equilibrium KDF is provided. As an application, the physical content of the
fluid 4-flow is analyzed and a comparison with the corresponding
non-relativistic limit of the kinetic solution is discussed. Finally, a
summary of the main achievements and concluding remarks are given in Section
11.

\section{The EM-tetrad basis}

For the covariant representation of the particle dynamics we use a formalism
based on the introduction of a tetrad of unit 4-vectors \cite{LL}. In the
present treatment the tetrad is associated with the properties of the EM
field through the Faraday tensor $F_{\mu \nu }=\partial _{\mu }A_{\nu
}-\partial _{\nu }A_{\mu }$. Notice that the EM-tetrad basis represents the
natural covariant generalization of the magnetic-related triad system formed
by the orthogonal right-handed unit 3-vectors $\left( \mathbf{e}_{1},\mathbf{%
e}_{2},\mathbf{e}_{3}\equiv \mathbf{b}\right) $ usually introduced in
non-relativistic treatments \cite{Cr2013}. In this section we summarize the
properties of such a frame; more details on the issue can be found in Refs.%
\cite{Bek,Bek1,bek3}.

The existence of the EM tetrad relies on the fact that any non-degenerate
antisymmetric tensor $F_{\mu \nu }$ has necessarily two orthogonal invariant
hyperplanes, which can be identified with the sets $\left( a^{\mu },b^{\mu
}\right) $ and $\left( c^{\mu },d^{\mu }\right) $, where $a^{\mu }$ and $%
\left( b^{\mu },c^{\mu },d^{\mu }\right) $ are respectively time-like and
space-like. One can show that each of the two invariant hyperplanes has a
single associated eigenvalue. In the following we denote with $H$ and $E$
the 4-scalar eigenvalues of $F_{\mu \nu }$. The physical meaning is that $H$
and $E$ coincide with the observable magnetic and electric field strengths
in the reference frame where the electric and the magnetic fields are
parallel. We shall assume that the relative velocity of such a reference
frame with respect to the observer reference frame is a prescribed
non-uniform 4-velocity $U_{R}^{\mu }\left( r\right) $. Its precise
definition depends on: 1)\ the observer reference frame; 2)\ the
representation of $F_{\mu \nu }$ in such a frame. Notice also that, by
construction, the EM-tetrad is defined in such a way that locally the metric
tensor $g_{\mu \nu }\left( r\right) \cong \eta _{\mu \nu }$ (condition of
local flatness, in turn based on the Einstein principle of equivalence), so
that in the EM-tetrad frame the controvariant and covariant basis $\left(
a^{\mu },b^{\mu },c^{\mu },d^{\mu }\right) $ and $\left( a_{\mu },b_{\mu
},c_{\mu },d_{\mu }\right) $ are related locally by means of $\eta _{\mu \nu
}$.

In the EM-tetrad frame the Faraday tensor can be represented as%
\begin{equation}
F_{\mu \nu }=H\left( c_{\nu }d_{\mu }-c_{\mu }d_{\nu }\right) +E\left(
b_{\mu }a_{\nu }-b_{\nu }a_{\mu }\right) .  \label{tetrad-fmunu}
\end{equation}%
Finally, the inverse of the Faraday tensor, i.e., $D^{\mu \nu }$, such that $%
F_{\mu \alpha }D^{\mu \nu }\equiv \delta _{\alpha }^{\nu }$, exists if $%
H,E\neq 0$, and is given by%
\begin{equation}
D_{\mu \nu }=-\frac{1}{H}\left( c_{\nu }d_{\mu }-c_{\mu }d_{\nu }\right) +%
\frac{1}{E}\left( b_{\mu }a_{\nu }-b_{\nu }a_{\mu }\right) .
\end{equation}%
Here it is assumed that $D_{\mu \nu }$ is defined and non-divergent in the
configuration domain of interest. In addition, we shall impose that both $E$
and $H$ satisfy the assumption of strong magnetization, in the sense that
everywhere in the system%
\begin{equation}
E^{2}-H^{2}<0,
\end{equation}%
while $H$ is parametrized in terms of a suitable dimensionless infinitesimal
4-scalar parameter $\varepsilon $ to be defined below (see Eq.(\ref{epsi})),
so that the following formal replacement is performed:%
\begin{equation}
H\rightarrow \frac{H}{\varepsilon },  \label{eehh}
\end{equation}%
where in the following $\frac{H}{\varepsilon }$ identifies the magnetic
field. Further, the metric tensor $g_{\mu \nu }\left( r\right) $ is left
invariant, i.e., formally $g_{\mu \nu }\left( r\right) $ is considered of $%
O\left( \varepsilon ^{0}\right) $. Such an asymptotic ordering for the
gravitational and magnetic fields is referred to as the \textit{%
strongly-magnetized ordering}.

As far as the electric field $E$ is concerned, we shall consider two
possible sub-orderings:

a) \textit{Strong electric field}, in which $E$ is replaced by%
\begin{equation}
E\rightarrow \frac{E}{\varepsilon },
\end{equation}%
which implies that the electric field cannot vanish.

b) \textit{Weak electric field}, in which $E$ is finite (with respect to $%
\varepsilon $) so that it is allowed to vanish locally or globally in the
configuration domain.

Let us notice that the previous asymptotic orderings correspond to
well-defined asymptotic conditions for the Faraday tensor which follow from
the tetrad representation (\ref{tetrad-fmunu}). In particular, the
components proportional to $H$ are always ordered as $1/O\left( \varepsilon
\right) $, while those proportional to $E$ are respectively $1/O\left(
\varepsilon \right) $ in case a) and of $O\left( \varepsilon ^{0}\right) $
in case b). This means that in the first case the 4-vector potential must be
replaced by%
\begin{equation}
A_{\mu }\rightarrow \frac{A_{\mu }}{\varepsilon }.  \label{tttt}
\end{equation}%
The same requirement applied also in case b), except for the time-component
which remains of $O\left( \varepsilon ^{0}\right) $.

We remark that the strongly-magnetized ordering considered in Refs.\cite%
{Bek,Bek1} actually refers to case a) only. However, we intend to show that
the GK theory can be extended also to encompass both regimes a) and b), so
to hold also when $E$ is permitted to vanish locally.

\subsection{Symmetry assumptions}

Here we want to deal with the general physical setting characterized by the
absence of spatial-symmetries. In general relativity this means that there
is no coordinate system in which at least the Faraday tensor admits an
ignorable space-like coordinate. In addition we shall require\ that in the
observer reference frame the corresponding time-like component $Q^{0}=ct$ of
the position 4-vector is ignorable for the single-particle Lagrangian $%
\mathcal{L}$, which in Lagrangian coordinates is given by%
\begin{equation}
\mathcal{L}\left( r,\frac{dr}{ds}\right) =\left( \frac{1}{2}g_{\mu \nu }%
\frac{dr^{\nu }}{ds}+q\frac{A_{\mu }}{\varepsilon }\right) \frac{dr^{\mu }}{%
ds},
\end{equation}%
where $q\equiv \frac{Ze}{M_{0}c^{2}}$ is the specific particle charge, with $%
M_{0}$ and $Ze$ being the rest-mass and charge of point-like particles (the
index of species is omitted from hereon). It follows that the conjugate
momentum $P_{0}=\frac{\partial \mathcal{L}}{\partial \left( \frac{dQ^{0}}{ds}%
\right) }$ is given by%
\begin{equation}
P_{0}=u_{0}+q\frac{A_{0}}{\varepsilon },  \label{p0}
\end{equation}%
where $u_{0}=g_{0\nu }\frac{dr^{\nu }}{ds}$. Hence, with respect to the
observer reference frame, $P_{0}$ is conserved provided $\frac{\partial 
\mathcal{L}}{\partial Q^{0}}=0$, namely both $g_{\mu \nu }$ and $A_{\mu }$
admit $Q^{0}$ as ignorable coordinate, i.e., in such a frame a stationarity
condition holds both for the gravitational and EM fields.

\section{Non-perturbative GK theory}

In this section we summarize the main results concerning the
non-perturbative formulation of the covariant GK theory developed in Ref.%
\cite{bek3}, which permits to establish the adiabatic conservation
properties of the relativistic particle magnetic moment, when
radiation-reaction effects are ignored \cite{EPJ1,EPJ2,EPJ3,EPJ4,EPJ5,EPJ6}.
In the present derivation of gyrokinetics\ charged particles are treated as
point-like and are assumed to belong to a magnetized plasma, in the sense
defined above. For definiteness, the background metric tensor $g_{\mu \nu
}\left( r\right) $ is considered a prescribed function of the position
4-vector $r^{\mu }$.

The relativistic formulation of GK theory is couched on the Hamilton
variational principle in synchronous form \cite{Bek1}. This is written in
terms of a super-abundant dynamical state, namely of the 8-dimensional
vector field $\mathbf{x}\equiv \left( r^{\mu },u^{\mu }\right) $.\textbf{\ }%
Notice that here only the extremal curves, which are determined by the
Euler-Lagrange equation associated with the same variational principle, are
identified with the particle 4-position and 4-velocity, $r^{\mu }(s)$ and $%
u^{\mu }(s)$, with $s$ being the particle proper-time. In particular, the
adoption of the synchronous principle implies that in the variational
principle the mass-shell condition%
\begin{equation}
u_{1}^{\prime \mu }u_{1\mu }^{\prime }=1  \label{mass-shell}
\end{equation}%
is actually satisfied only by the extremal curves.\textbf{\ }As such, the
action functional can then be represented in terms of arbitrary hybrid
(i.e., non-canonical) variables. This permits the introduction of an
extended phase-state transformation\textbf{\ }which is realized by a local
diffeomorphism of the form \cite{Bek1,Bek}%
\begin{equation}
\mathbf{x}\equiv \left( r^{\mu },u^{\mu }\right) \leftrightarrow \mathbf{z}%
^{\prime }\equiv \left( \mathbf{y}^{\prime },\phi ^{\prime }\right) .
\label{87}
\end{equation}%
Here $\phi ^{\prime }$ is a suitable gyrophase angle to be suitably defined
(see below)\ while $\mathbf{z}^{\prime }$ is referred to as the GK state,
with $\mathbf{y}^{\prime }$ being a suitable 7-component vector. Notice that
locality implies that it must be possible to represent $\mathbf{z}^{\prime }$
in the form $\mathbf{z}^{\prime }=\mathbf{z}^{\prime }\left( \mathbf{x}%
\right) $, as well as its inverse $\mathbf{x}=\mathbf{x}\left( \mathbf{z}%
^{\prime }\right) $. The GK state $\mathbf{z}^{\prime }$ is constructed in
such a way that its equations of motion are gyrophase independent, namely $%
\frac{d}{ds}\mathbf{z}^{\prime }\equiv \mathbf{F}(\mathbf{y}^{\prime },s)$,
where $\mathbf{F}$ is a suitable vector field.

A non-perturbative covariant GK theory is established by introducing first
an extended local transformation of the type%
\begin{eqnarray}
r^{\mu } &=&r^{\prime \mu }+\rho _{1}^{\prime \mu },  \label{1} \\
u^{\mu } &=&u^{\prime \mu }\oplus \nu _{1}^{\prime \mu },  \label{2}
\end{eqnarray}%
denoted as extended guiding-center transformation, where $\rho _{1}^{\prime
\mu }=\rho _{1}^{\prime \mu }\left( r^{\prime \mu },u^{\prime \mu }\right) $
and $\nu _{1}^{\prime \mu }=\nu _{1}^{\prime \mu }\left( r^{\prime \mu
},u^{\prime \mu }\right) $ are suitably prescribed in terms of $\left(
r^{\prime \mu },u^{\prime \mu }\right) $. Here the notation is analogous to
Refs.\cite{Bek1,Bek,bek3}. In particular, $r^{\prime \mu }$ is the
guiding-center position 4-vector, with primed quantities denoting dynamical
variables which are evaluated at $r^{\prime \mu }$. Thus, $\rho _{1}^{\prime
\mu }$ is referred to as the relativistic Larmor 4-vector, while both $%
u^{\mu }$ and $u^{\prime \mu }$ are by construction 4-velocities, so that $%
u^{\mu }u_{\mu }=u^{\prime \mu }u_{\mu }^{\prime }=1$, with $\oplus $
denoting the relativistic 4-velocity composition law%
\begin{equation}
u^{\prime \mu }\oplus \nu _{1}^{\prime \mu }=\frac{u^{\prime \mu }+\nu
_{1}^{\prime \mu }}{\sqrt{1+\nu _{1\mu }^{\prime }\nu _{1}^{\prime \mu
}+2u_{\mu }^{\prime }\nu _{1}^{\prime \mu }}}.  \label{comp-law}
\end{equation}%
with $\nu _{1}^{\prime \mu }$ to be defined so that the condition $\nu
_{1}^{\prime \mu }\nu _{1\mu }^{\prime }\neq 1$ holds. Notice that by
construction: 1) $\nu _{1}^{\prime \mu }$ is not a 4-velocity; 2) $%
u_{1}^{\prime \mu }\equiv $ $u^{\prime \mu }\oplus \nu _{1}^{\prime \mu }$
is necessarily a 4-velocity provided $u^{\prime \mu }$ is so; as a
consequence in such a case $u_{1}^{\prime \mu }$ satisfies identically the
mass-shell condition (\ref{mass-shell}).

Notice that the adoption of the transformation law (\ref{comp-law}) is a
basic physical prerequisite needed to warrant that the extremal curve $%
u^{\prime \mu }(s)$ is also a physically realizable 4-velocity. A\ further
constraint, however, must be taken into account, which affects also the
variational curves. This is provided by the requirement that the
perturbations $\rho _{1}^{\prime \mu }$\ and $\nu _{1}^{\prime \mu }$\ are
suitably related.\textbf{\ }This means that the guiding-center
transformation (\ref{1}) and (\ref{2}) are required to fulfill the equation%
\begin{equation}
\frac{d}{ds}\left( r^{\prime \mu }+\rho _{1}^{\prime \mu }\right) =u^{\prime
\mu }\oplus \nu _{1}^{\prime \mu },  \label{constraint-extremal}
\end{equation}%
which relates the transformed physical velocity to the rate of change of the
displacement vector $r^{\prime \mu }+\rho _{1}^{\prime \mu }$. Therefore,
the introduction of a non-vanishing Larmor-radius displacement $\rho
_{1}^{\prime \mu }$ requires also the general introduction of a
corresponding non-vanishing perturbation $\nu _{1}^{\prime \mu }$ in the
physical 4-velocity. \textbf{This} physical requirement is mandatory and
cannot be ignored.

Let us now project $u^{\prime \mu }$ along the EM-tetrad $\left( a^{\prime
\mu },b^{\prime \mu },c^{\prime \mu },d^{\prime \mu }\right) $ evaluated at
the guiding-center position (guiding-center EM-tetrad). Then, the following
representation is assumed:%
\begin{equation}
u^{\prime \mu }\equiv u_{0}^{\prime }a^{\prime \mu }+u_{\parallel }^{\prime
}b^{\prime \mu }+w^{\prime }\left[ c^{\prime \mu }\cos \phi ^{\prime
}+d^{\prime \mu }\sin \phi ^{\prime }\right] ,  \label{u-tetrade}
\end{equation}%
where%
\begin{equation}
u_{0}^{\prime }=\sqrt{1+u_{\parallel }^{\prime 2}+w^{\prime 2}},  \label{u00}
\end{equation}%
and the gyrophase angle $\phi ^{\prime }$ is introduced. This implies the
definition%
\begin{equation}
\phi ^{\prime }=\arctan \left[ \frac{u^{\prime \mu }d_{\mu }^{\prime }}{%
u^{\prime \mu }c_{\mu }^{\prime }}\right] ,
\end{equation}%
where by construction $\phi ^{\prime }$ is a 4-scalar. As a consequence of
the definition (\ref{u-tetrade}) also the components\ $u_{0}^{\prime }$, $%
u_{\parallel }^{\prime }$ and $w^{\prime }$ are manifestly 4-scalars.
Analogous decompositions follow also for $\rho _{1}^{\prime \mu }$ and $\nu
_{1}^{\prime \mu }$, which are expected to have non-vanishing components
along all the directions of the EM basis.

Thus, denoting by $\left\langle h(\mathbf{z}^{\prime })\right\rangle _{\phi
^{\prime }}$ the gyrophase-average operator acting on a function $h(\mathbf{z%
}^{\prime })$, the quantity $\left[ h(\mathbf{z}^{\prime })\right] ^{\sim
}\equiv h(\mathbf{z}^{\prime })-\left\langle h(\mathbf{z}^{\prime
})\right\rangle _{\phi ^{\prime }}$ is referred to as the oscillatory part
of $h$. The definition of the gyrophase-average operator is given in Ref.%
\cite{bek3} and depends on the identification of the GK variables. We remark
that here such operator does not change the tensorial property of $h(\mathbf{%
z}^{\prime })$, hence it is a 4-scalar operator. We notice that in the
present theory only $\rho _{1}^{\prime \mu }$ is assumed to be purely
oscillatory by construction, so that $\left\langle \rho _{1}^{\prime \mu
}\right\rangle _{\phi ^{\prime }}=0$, while for $\nu _{1}^{\prime \mu }$ we
must impose $\left\langle \nu _{1}^{\prime \mu }\right\rangle _{\phi
^{\prime }}\neq 0$ (see also discussion below). Similarly, $r^{\prime \mu }$%
, $u_{0}^{\prime }$, $u_{\parallel }^{\prime }$ and $w^{\prime }$ are
gyrophase-independent. In the following, the GK state is prescribed in terms
of the gyrophase angle $\phi ^{\prime }$ by imposing Eq.(\ref{87}), while
the remaining 7-component vector $\mathbf{y}^{\prime }$ is taken to be
generally a function of the remaining six independent variables. For
example, the one which is adopted in the following is $\mathbf{y}^{\prime }=%
\mathbf{y}^{\prime }\left( r^{\prime \mu },u_{\parallel }^{\prime },\mu
^{\prime }\right) $, where $\mu ^{\prime }$ is the leading-order
contribution to the magnetic moment $m^{\prime }$ (see below).

The derivation of the GK equations of motion and of the related conservation
laws follow standard procedures in the framework of variational Lagrangian
approach. The detailed steps in this regard are illustrated in Ref.\cite%
{bek3}, where both the Lagrangian differential 1-form $\delta \mathcal{L}$
and the corresponding gyrophase-independent transformed differential form $%
\delta \mathcal{L}_{1}$\textbf{\ }are introduced. A basic consequence is
that in terms of $\delta \mathcal{L}_{1}$ it is possible to determine an
exact and non-perturbative representation of the particle magnetic moment $%
m^{\prime }$. Following Ref.\cite{bek3}, this is found to be%
\begin{equation}
m^{\prime }=\left\langle \frac{\partial \rho _{1}^{\prime \mu }}{\partial
\phi ^{\prime }}\left[ \left( u_{\mu }^{\prime }\oplus \nu _{1\mu }^{\prime
}\right) +qA_{\mu }\right] \right\rangle _{\phi ^{\prime }},  \label{m-exact}
\end{equation}%
which represents the covariant generalization of the analogous
non-relativistic result obtained in Ref.\cite{Cr2013}.

The representation (\ref{m-exact}) given above of course holds only provided
the two transformations (\ref{87}) and (\ref{1})-(\ref{2}) actually exist,
i.e. are invertible. A number of important qualitative features must be
pointed out. First, by construction $m^{\prime }$ is a 4-scalar, and
therefore it is an observable with a unique local value, which is also
frame-independent. Second, in analogy with the non-relativistic theory, $%
m^{\prime }$ can be interpreted as a canonical momentum. Third, in this
expression $A_{\mu }$ depends explicitly on the gyrophase via Eq.(\ref{1}).
Finally, it is obvious that the rhs of Eq.(\ref{m-exact}) exhibits explicit
dependences in terms of all the independent components of the 4-velocity $%
u^{\prime \mu }$, i.e., $u_{\parallel }^{\prime }$ and $w^{\prime }$.

It is important to stress that, at this stage, the transformations relating $%
\mathbf{x}$, $\mathbf{z}^{\prime }$ and $\mathbf{y}^{\prime }$ are in
principle finite and non-perturbative, so that they are not based on any
asymptotic expansion for the guiding-center transformation (\ref{1})-(\ref{2}%
) and/or the construction of the GK Lagrangian differential form $\delta 
\mathcal{L}_{1}$. As a fundamental physical implication, in the subset of
phase-space in which the GK transformation (\ref{87}) is defined, provided $%
\rho _{1}^{\prime \mu }$ and $v_{1}^{\prime \mu }$ are uniquely determined,
the relativistic magnetic moment $m^{\prime }$ is an integral of motion.

\section{Perturbative GK theory}

In this section we determine a covariant perturbative theory appropriate for
the analytical asymptotic treatment of the exact GK Lagrangian formulation
established above. Standard perturbative methods can be adopted for this
purpose \cite{Bek1,Bek,lit81,hahm}, which rely on the so-called
Larmor-radius expansion. To permit a perspicuous comparison with the
non-relativistic treatment, the direct construction method is adopted for
the perturbative expansion, as in Ref.\cite{Cr2013}. In the context of a
covariant formulation of a perturbative theory, the relevant expansion
parameters are necessarily 4-scalars. For GK theory this is identified with
the dimensionless Larmor-radius parameter, namely the frame-invariant ratio%
\begin{equation}
\varepsilon \equiv \frac{r_{L}}{L}\ll 1,  \label{epsi}
\end{equation}%
to be considered as an infinitesimal. Here $r_{L}$ is the 4-scalar $%
r_{L}\equiv \sqrt{\rho _{1}^{\prime \mu }\rho _{1\mu }^{\prime }}$, with $%
\rho _{1}^{\prime \mu }$ being the Larmor-radius 4-vector which enters in
the definition of the guiding-center 4-position transformation (\ref{1}).
Furthermore, $L$ is the 4-scalar defined as%
\begin{equation}
\frac{1}{L^{2}}\equiv \sup \left[ \frac{1}{\Lambda ^{2}}\partial _{\mu
}\Lambda \partial ^{\mu }\Lambda \right] ,
\end{equation}%
with $\Lambda \left( r\right) $ denoting\ here either the set of scalar
fields $E^{2}-H^{2}$ or the Kretschmann invariant $K\equiv R^{\alpha \beta
\gamma \delta }R_{\alpha \beta \gamma \delta }$ associated with the metric
tensor \cite{grav}. From the definition (\ref{epsi}) it follows that $\rho
_{1}^{\prime \mu }$ must be considered infinitesimal, so that one can
introduce the formal representation%
\begin{equation}
\rho _{1}^{\prime \mu }\rightarrow \varepsilon \rho _{1}^{\prime \mu }
\end{equation}%
in the guiding-center transformation (\ref{1}). In complete analogy, in Eq.(%
\ref{2}) $\nu _{1}^{\prime \mu }$ is treated as a perturbation, namely
requiring also%
\begin{equation}
\nu _{1}^{\prime \mu }\rightarrow \varepsilon \nu _{1}^{\prime \mu }.
\end{equation}%
Instead, in the two guiding-center transformations (\ref{1}) and (\ref{2})
the 4-vectors $r^{\prime \mu }$ and $u^{\prime \mu }$ are considered to be $%
O\left( \varepsilon ^{0}\right) $.

We now invoke the assumption of analyticity for $\varepsilon \rho
_{1}^{\prime \mu }$ and $\varepsilon \nu _{1}^{\prime \mu }$ in terms of $%
\varepsilon $, so that they can be represented in terms of the power series%
\begin{eqnarray}
\varepsilon \rho _{1}^{\prime \mu } &=&\varepsilon r_{1}^{\prime \mu
}+\varepsilon ^{2}r_{2}^{\prime \mu }+..., \\
\varepsilon \nu _{1}^{\prime \mu } &=&\varepsilon v_{1}^{\prime \mu
}+\varepsilon ^{2}v_{2}^{\prime \mu }+....
\end{eqnarray}%
Furthermore, here for greater generality the 4-vector potential is ordered
according to the strong electric field ordering, namely $\frac{A_{\mu }}{%
\varepsilon }\equiv \frac{A_{\mu }\left( r\right) }{\varepsilon }$, with $%
A_{\mu }\left( r\right) $\ being required to satisfy as usual the Lorentz
gauge. We shall assume in particular that both $\frac{A_{\mu }\left(
r\right) }{\varepsilon }$ and the metric tensor $g_{\mu \nu }=g_{\mu \nu
}(r) $\ are analytic functions of the 4-position $r^{\mu }$. This means that
both tensors can have in principle fast dependences, namely of $O\left(
\varepsilon ^{0}\right) $ in terms of $r^{\mu }$. Then by construction $%
\frac{A_{\mu }\left( r\right) }{\varepsilon }$ admits an
asymptotically-convergent power series Taylor expansion of the form%
\begin{eqnarray}
&&\left. \frac{A_{\mu }}{\varepsilon }=\frac{A_{\mu }^{\prime }}{\varepsilon 
}+r_{1}^{\prime \nu }\frac{\partial A_{\mu }^{\prime }}{\partial r^{\prime
\nu }}\right.  \notag \\
&&+\varepsilon \left( r_{2}^{\prime \nu }\frac{\partial A_{\mu }^{\prime }}{%
\partial r^{\prime \nu }}+\frac{1}{2}r_{1}^{\prime \nu }r_{1}^{\prime
\varsigma }\frac{\partial ^{2}A_{\mu }^{\prime }}{\partial r^{\prime \nu
}\partial r^{\prime \varsigma }}\right) +O\left( \varepsilon ^{2}\right) .
\label{fin}
\end{eqnarray}%
In the following we shall consider a perturbative guiding-center
transformation of the form%
\begin{eqnarray}
r^{\mu } &=&r^{\prime \mu }+\varepsilon r_{1}^{\prime \mu }+\varepsilon
^{2}r_{2}^{\prime \mu },  \label{1-bis} \\
u^{\mu } &=&u^{\prime \mu }+\varepsilon V^{\prime \mu },  \label{2-bis}
\end{eqnarray}%
where, from the 4-velocity composition law (\ref{comp-law}), the 4-vector $%
V^{\prime \mu }$ is given by%
\begin{equation}
\varepsilon V^{\prime \mu }\equiv \varepsilon v_{1}^{\prime \mu
}-\varepsilon u^{\prime \alpha }v_{1\alpha }^{\prime }u^{\prime \mu }.
\label{v-grande}
\end{equation}

These expansions allow one to express the constraint (\ref%
{constraint-extremal}) holding on the extremal variables as%
\begin{equation}
\frac{d}{ds}\left[ r^{\prime \mu }+\varepsilon r_{1}^{\prime \mu
}+\varepsilon ^{2}r_{2}^{\prime \mu }\right] =u^{\prime \mu }+\varepsilon
V^{\prime \mu },
\end{equation}%
which, correct to first order, yields the two extremal perturbative equations%
\begin{equation}
\frac{dr^{\prime \mu }}{ds}+\overset{\cdot }{\phi }\frac{\partial
r_{1}^{\prime \mu }}{\partial \phi }=u^{\prime \mu },  \label{ext-1}
\end{equation}%
\begin{equation}
(\frac{dr_{1}^{\prime \mu }}{ds})_{\phi ^{\prime }}+\overset{\cdot }{\phi }%
\frac{\partial r_{2}^{\prime \mu }}{\partial \phi }=V^{\prime \mu }.
\label{ext-2}
\end{equation}%
In the last equation the lhs is purely oscillatory, so that also $V^{\prime
\mu }$ must satisfy the constraint $\left\langle V^{\prime \mu
}\right\rangle _{\phi ^{\prime }}=0$. As a consequence, from Eq.(\ref%
{v-grande}) it follows that the rhs of Eq.(\ref{ext-2}) must satisfy the
further constraint equation%
\begin{eqnarray}
&&\left. \left\langle v_{1}^{\prime \mu }\right\rangle _{\phi ^{\prime
}}=\left\langle u^{\prime \alpha }v_{1\alpha }^{\prime }u^{\prime \mu
}\right\rangle _{\phi ^{\prime }}\right.  \notag \\
&&\left. =\left\langle v_{1\alpha }^{\prime }\right\rangle _{\phi ^{\prime
}}\left\langle u^{\prime \alpha }u^{\prime \mu }\right\rangle _{\phi
^{\prime }}+\left\langle u^{\prime \alpha }\left[ v_{1\alpha }^{\prime }%
\right] ^{\sim }u^{\prime \mu }\right\rangle _{\phi ^{\prime }},\right.
\end{eqnarray}%
which is an equation for $\left\langle v_{1}^{\prime \mu }\right\rangle
_{\phi ^{\prime }}$ only, with $\left[ v_{1\alpha }^{\prime }\right] ^{\sim
} $ to be considered prescribed in terms of $V^{\prime \mu }$. We conclude
noticing that in principle all these equations must be taken into account
below because they provide the physical constraints on the guiding-center
transformation. Nevertheless, the explicit determination of $\left\langle
v_{1}^{\prime \mu }\right\rangle _{\phi ^{\prime }}$ is actually irrelevant
because only $V^{\prime \mu }=\left[ V^{\prime \mu }\right] ^{\sim }$
matters.

\section{Lagrangian differential form in GK variables}

Let us now represent the fundamental Lagrangian differential form in terms
of the truncated guiding-center transformations (\ref{1-bis})-(\ref{2-bis})
and the truncated perturbative expansion (\ref{fin}) of the EM 4-potential.
As specified above, we now select the set of seven independent GK variables
to be $\mathbf{z}^{\prime }=\left( r^{\prime \mu },u_{\parallel }^{\prime
},\mu ^{\prime },\phi ^{\prime }\right) $, where $\mu ^{\prime }$ denotes
the leading-order contribution of the guiding-center magnetic moment.\textbf{%
\ }This implies that the differential $dr^{\mu }$ which enters the
Lagrangian $\delta \mathcal{L}$ can be formally expressed as%
\begin{equation}
dr^{\mu }=(dr^{\mu })_{\phi ^{\prime }}+\frac{1}{\varepsilon }\frac{\partial
r^{\mu }}{\partial \phi ^{\prime }}d\phi ^{\prime },
\end{equation}%
where we have singled out the partial derivative with respect to $\phi
^{\prime }$, so that $(dr^{\mu })_{\phi ^{\prime }}$ means the differential
of $r^{\mu }$ with respect to all other GK variables of the set $\mathbf{z}%
^{\prime }$ keeping $\phi ^{\prime }$ constant. Notice that the factor $%
1/\varepsilon $ is introduced to denote the fast dependence that
characterizes the Larmor rotation \cite{Bek1}. Substituting the expansion (%
\ref{1-bis}) then gives, neglecting corrections of $O(\varepsilon ^{3})$,%
\begin{eqnarray}
dr^{\mu } &=&dr^{\prime \mu }+\frac{\partial r_{1}^{\prime \mu }}{\partial
\phi ^{\prime }}d\phi ^{\prime }+\varepsilon (dr_{1}^{\prime \mu })_{\phi
^{\prime }}  \notag \\
&&+\varepsilon \frac{\partial r_{2}^{\prime \mu }}{\partial \phi ^{\prime }}%
d\phi ^{\prime }+\varepsilon ^{2}(dr_{2}^{\prime \mu })_{\phi ^{\prime }}.
\label{dr}
\end{eqnarray}%
Assuming for $r_{1}^{\prime \mu }$ a general dependence of the type $%
r_{1}^{\prime \mu }=r_{1}^{\prime \mu }\left( r^{\prime },\xi ,\mu ^{\prime
},\phi ^{\prime }\right) $ it follows that:%
\begin{equation}
\varepsilon (dr_{1}^{\prime \mu })_{\phi ^{\prime }}=\varepsilon \frac{%
\partial r_{1}^{\prime \mu }}{\partial r^{\prime \alpha }}dr^{\prime \alpha
}+\varepsilon \frac{\partial r_{1}^{\prime \mu }}{\partial u_{\parallel
}^{\prime }}du_{\parallel }^{\prime }+\varepsilon \frac{\partial
r_{1}^{\prime \mu }}{\partial \mu ^{\prime }}d\mu ^{\prime },  \label{dr-fi}
\end{equation}%
with a similar expression holding also for $(dr_{2}^{\prime \mu })_{\phi
^{\prime }}$.

Substituting these expressions in $\delta \mathcal{L}$ yields the
gauge-modified functional $\delta \mathcal{L}^{\prime }$ defined as%
\begin{equation}
\delta \mathcal{L}^{\prime }\equiv \delta \mathcal{L-}dG-\varepsilon
dR-\varepsilon ^{2}dR_{1},
\end{equation}%
where $G$, $R$ and $R_{1}$ are scalar gauge functions. In particular, the
dynamical gauge $G$ is given by%
\begin{equation}
G\equiv \frac{q}{2}\left( A_{\mu }+A_{\mu }^{\prime }\right) \left(
r_{1}^{\prime \mu }+\varepsilon r_{2}^{\prime \mu }\right) ,
\end{equation}%
while $R$ and $R_{1}$ are still undetermined at this stage (notice however
that they are actually determined uniquely by the perturbative approach, see
below).\textbf{\ }As shown in Ref.\cite{Bek} this represents a convenient
choice of the gauge $G$\ because it permits to display the explicit
dependence of the Larmor radius in terms of the observable Faraday tensor.
After some algebra, the resulting expression for $\delta \mathcal{L}^{\prime
}$ becomes%
\begin{eqnarray}
\delta \mathcal{L}^{\prime } &=&\left[ \left( \Gamma _{r^{\prime \mu
}}\right) _{\mu }-\varepsilon \frac{\partial R}{\partial r^{\prime \mu }}%
\right] dr^{\prime \mu }+\left[ \Gamma _{u_{\parallel }^{\prime
}}-\varepsilon \frac{\partial R}{\partial u_{\parallel }^{\prime }}\right]
du_{\parallel }^{\prime }  \notag \\
&&+\left[ \Gamma _{\phi ^{\prime }}-\frac{\partial R}{\partial \phi ^{\prime
}}-\varepsilon \frac{\partial R_{1}}{\partial \phi ^{\prime }}\right] d\phi
^{\prime },  \label{dg'}
\end{eqnarray}%
where the coefficients $\left( \Gamma _{r^{\prime \mu }}\right) _{\mu }$, $%
\Gamma _{u_{\parallel }^{\prime }}$ and $\Gamma _{\phi ^{\prime }}$ are
respectively%
\begin{eqnarray}
\left( \Gamma _{r^{\prime \mu }}\right) _{\mu } &\equiv &\frac{1}{%
\varepsilon }qA_{\mu }^{\prime }+u_{\mu }^{\prime }+qr_{1}^{\prime \nu
}F_{\nu \mu }^{\prime }  \notag \\
&&+\varepsilon V_{\mu }^{\prime }+\varepsilon qr_{2}^{\prime \nu }F_{\nu \mu
}^{\prime }+\varepsilon q\frac{1}{2}r_{1}^{\prime \nu }r_{1}^{\prime
\varsigma }\frac{\partial F_{\varsigma \mu }^{\prime }}{\partial r^{\prime
\nu }}  \notag \\
&&+\varepsilon \frac{\partial r_{1}^{\prime \beta }}{\partial r^{\prime \mu }%
}\left( u_{\beta }^{\prime }+\frac{1}{2}qr_{1}^{\prime \nu }F_{\nu \beta
}^{\prime }\right) ,  \label{gamma-r}
\end{eqnarray}%
\begin{equation}
\Gamma _{u_{\parallel }^{\prime }}\equiv \varepsilon \frac{\partial
r_{1}^{\prime \mu }}{\partial u_{\parallel }^{\prime }}\left( u_{\mu
}^{\prime }+\frac{1}{2}qr_{1}^{\prime \nu }F_{\nu \mu }^{\prime }\right) ,
\label{gamma-u}
\end{equation}%
and%
\begin{eqnarray}
\Gamma _{\phi ^{\prime }} &\equiv &\frac{\partial r_{1}^{\prime \mu }}{%
\partial \phi ^{\prime }}\left( u_{\mu }^{\prime }-\frac{q}{2}F_{\mu \nu
}^{\prime }r_{1}^{\prime \nu }\right)  \notag \\
&&+\varepsilon \frac{\partial r_{2}^{\prime \mu }}{\partial \phi ^{\prime }}%
u_{\mu }^{\prime }+\varepsilon \frac{\partial r_{1}^{\prime \mu }}{\partial
\phi ^{\prime }}V_{\mu }^{\prime }+\varepsilon \frac{q}{2}r_{1}^{\prime \nu }%
\frac{\partial r_{2}^{\prime \mu }}{\partial \phi ^{\prime }}F_{\nu \mu
}^{\prime }  \notag \\
&&+\varepsilon \frac{q}{2}r_{2}^{\prime \nu }\frac{\partial r_{1}^{\prime
\mu }}{\partial \phi ^{\prime }}F_{\nu \mu }^{\prime }+\varepsilon \frac{q}{2%
}r_{1}^{\prime \varsigma }r_{1}^{\prime \nu }\frac{\partial ^{2}A_{\mu
}^{\prime }}{\partial r^{\prime \nu }\partial r^{\prime \varsigma }}\frac{%
\partial r_{1}^{\prime \mu }}{\partial \phi ^{\prime }}.
\end{eqnarray}

The GK Lagrangian $\delta \mathcal{L}_{1}$ is then obtained by imposing the
requirement that the oscillatory part of $\delta \mathcal{L}^{\prime }$
vanishes identically. Indeed, the oscillatory parts are also periodic in
gyrophase.\textbf{\ }Therefore, it is always possible to impose the validity
of the following constraint conditions:%
\begin{eqnarray}
\left[ \left( \Gamma _{r^{\prime \mu }}\right) _{\mu }-\varepsilon \frac{%
\partial R}{\partial r^{\prime \mu }}\right] ^{\sim } &=&0,  \label{cc-1} \\
\left[ \Gamma _{u_{\parallel }^{\prime }}-\varepsilon \frac{\partial R}{%
\partial u_{\parallel }^{\prime }}\right] ^{\sim } &=&0,  \label{cc-2} \\
\left[ \Gamma _{\phi ^{\prime }}-\frac{\partial R}{\partial \phi ^{\prime }}%
-\varepsilon \frac{\partial R_{1}}{\partial \phi ^{\prime }}\right] ^{\sim }
&=&0,  \label{cc-3}
\end{eqnarray}%
which are demanded in order to realize the GK transformation. Provided Eqs.(%
\ref{cc-1})-(\ref{cc-3}) are identically satisfied, the Lagrangian
differential form $\delta \mathcal{L}^{\prime }$ becomes
gyrophase-independent. As a result, one obtains the GK Lagrangian function
expressed as $\delta \mathcal{L}_{1}\equiv \left\langle \delta \mathcal{L}%
^{\prime }\right\rangle _{\phi ^{\prime }}$, so that in this approximation%
\begin{eqnarray}
\delta \mathcal{L}_{1} &=&\left\langle \left( \Gamma _{r^{\prime \mu
}}\right) _{\mu }-\varepsilon \frac{\partial R}{\partial r^{\prime \mu }}%
\right\rangle _{\phi ^{\prime }}dr^{\prime \mu }  \notag \\
&&+\left\langle \Gamma _{u_{\parallel }^{\prime }}-\varepsilon \frac{%
\partial R}{\partial u_{\parallel }^{\prime }}\right\rangle _{\phi ^{\prime
}}du_{\parallel }^{\prime }  \notag \\
&&+\left\langle \Gamma _{\phi ^{\prime }}-\frac{\partial R}{\partial \phi
^{\prime }}-\varepsilon \frac{\partial R_{1}}{\partial \phi ^{\prime }}%
\right\rangle _{\phi ^{\prime }}d\phi ^{\prime }.  \label{GK-Lag}
\end{eqnarray}

In particular, the gyro-averaged coefficient of $d\phi ^{\prime }$
determines the expression of the magnetic moment correct to second-order in
the Larmor-radius expansion. Indeed, recalling that by construction $\frac{%
\partial R}{\partial \phi ^{\prime }}$ and $\frac{\partial R_{1}}{\partial
\phi ^{\prime }}$ are purely oscillatory, this yields 
\begin{equation}
\left\langle \Gamma _{\phi ^{\prime }}\right\rangle _{\phi ^{\prime }}=\mu
^{\prime }+\varepsilon \mu _{1}^{\prime }=m^{\prime }\left[ 1+O\left(
\varepsilon ^{2}\right) \right] .  \label{m-perturb}
\end{equation}%
As a last step, the theory must be completed by imposing the simultaneous
validity of the constraint equations (\ref{ext-1}) and (\ref{ext-2})
applying for the extremal dynamical variables and providing a relationship
for $r_{1}^{\prime \mu }$, $r_{2}^{\prime \mu }$ and $v_{1}^{\prime \mu }$.

At this point one can proceed solving the constraint equations at each order
in $\varepsilon $. In particular, the $O\left( \varepsilon ^{0}\right) $
perturbative solution can be found in Ref.\cite{bek3}. This provides an
explicit solution for the leading-order Larmor radius $r_{1}^{\prime \nu }$
as%
\begin{equation}
r_{1}^{\prime \nu }=-\frac{w^{\prime }}{qH^{\prime }}(d^{\prime \nu }\sin
\phi ^{\prime }-c^{\prime }{}^{\nu }\cos \phi ^{\prime }).
\label{larmor-radius-order1}
\end{equation}%
Notice that this expression is consistent with the constraint equation (\ref%
{ext-1}) upon identifying the leading-order relativistic Larmor frequency $%
\overset{\cdot }{\phi }=-\Omega ^{\prime }$, where $\Omega ^{\prime
}=qH^{\prime }$\ is denoted as effective Larmor frequency, which has the
dimension $\left[ L^{-1}\right] $\ and is evaluated at 4-position $r^{\prime
\mu }$. Then one finds that $\Gamma _{u_{\parallel }^{\prime }}=0$, while $%
R=R(\mu ^{\prime },\phi ^{\prime })$, namely it does not depend explicitly
on $r^{\prime \mu }$. As a result Eq.(\ref{cc-2}) is identically satisfied
at this order. Finally, the $O\left( \varepsilon ^{0}\right) $\ magnetic
moment $\mu ^{\prime }$\ takes the customary representation as%
\begin{equation}
\mu ^{\prime }=\left\langle \frac{\partial r_{1}^{\prime \mu }}{\partial
\phi ^{\prime }}\left( u_{\mu }^{\prime }-\frac{1}{2}qF_{\mu \nu }^{\prime
}r_{1}^{\prime \nu }\right) \right\rangle _{\phi ^{\prime }}=\frac{w^{\prime
2}}{2qH^{\prime }}.  \label{mu-primo}
\end{equation}

\section{Perturbative solution: $O\left( \protect\varepsilon \right) $}

In this section we proceed evaluating the perturbative GK theory to $O\left(
\varepsilon \right) $ in the Larmor-radius expansion, with the purpose of
determining the velocity dependences carried by the first-order magnetic
moment $\mu _{1}^{\prime }$. In view of the conclusions following from the
leading-order solution, the solubility conditions to be considered are given
by Eqs.(\ref{ext-2}), (\ref{cc-1}) and (\ref{cc-3}), where the $O\left(
\varepsilon \right) $-terms must be retained now.

We start by Eq.(\ref{cc-3}), which determines uniquely the gauge function $%
R_{1}$ as%
\begin{eqnarray}
&&\left[ \frac{\partial r_{2}^{\prime \mu }}{\partial \phi ^{\prime }}u_{\mu
}^{\prime }+\frac{\partial r_{1}^{\prime \mu }}{\partial \phi ^{\prime }}%
V_{\mu }^{\prime }+\frac{1}{2}qr_{1}^{\prime \nu }\frac{\partial
r_{2}^{\prime \mu }}{\partial \phi ^{\prime }}F_{\nu \mu }^{\prime }+\frac{q%
}{2}r_{2}^{\prime \nu }\frac{\partial r_{1}^{\prime \mu }}{\partial \phi
^{\prime }}F_{\nu \mu }^{\prime }\right] ^{\sim }  \notag \\
&&+\left[ \frac{q}{2}r_{1}^{\prime \varsigma }r_{1}^{\prime \nu }\frac{%
\partial ^{2}A_{\mu }^{\prime }}{\partial r^{\prime \nu }\partial r^{\prime
\varsigma }}\frac{\partial r_{1}^{\prime \mu }}{\partial \phi ^{\prime }}%
\right] ^{\sim }=\frac{\partial R_{1}}{\partial \phi ^{\prime }}.
\end{eqnarray}%
The next constraint equation (\ref{cc-1}) yields%
\begin{eqnarray}
&&\left[ V_{\mu }^{\prime }+qr_{2}^{\prime \nu }F_{\nu \mu }^{\prime }+q%
\frac{1}{2}r_{1}^{\prime \nu }r_{1}^{\prime \varsigma }\frac{\partial
F_{\varsigma \mu }^{\prime }}{\partial r^{\prime \nu }}\right] ^{\sim } 
\notag \\
&&\left. +\left[ \frac{\partial r_{1}^{\prime \beta }}{\partial r^{\prime
\mu }}\left( u_{\beta }^{\prime }-\frac{1}{2}qF_{\beta \nu }^{\prime
}r_{1}^{\prime \nu }\right) \right] ^{\sim }=0.\right.  \label{dr-tilda-3}
\end{eqnarray}%
This\ tensorial equation contains two unknown variables, respectively $%
r_{2}^{\prime \nu }$ and $V_{\mu }^{\prime }$ (or equivalently $%
v_{1}^{\prime \mu }$) and it must be coupled to Eq.(\ref{ext-2}). Since, as
indicated above, $r_{1}^{\prime \mu }$ does not depend on $u_{\parallel
}^{\prime }$, while by construction $\mu ^{\prime }$ is an adiabatic
invariant of $O\left( \varepsilon \right) $, so that $d\mu ^{\prime }\sim
O\left( \varepsilon \right) $, then from Eq.(\ref{ext-2}) it follows that $%
V_{\mu }^{\prime }$ takes the form%
\begin{equation}
V^{\prime \mu }=\left( u_{0}^{\prime }a^{\prime \alpha }+u_{\parallel
}^{\prime }b^{\prime \alpha }\right) \frac{\partial r_{1}^{\prime \mu }}{%
\partial r^{\prime \alpha }}+\overset{\cdot }{\phi }\frac{\partial
r_{2}^{\prime \mu }}{\partial \phi }.  \label{v-grande-bis}
\end{equation}%
As a consequence the constraint equation (\ref{dr-tilda-3}) reduces to a PDE
for the still undetermined 4-vector $r_{2}^{\prime \mu }$:%
\begin{equation}
\overset{\cdot }{\phi ^{\prime }}\frac{\partial r_{2\mu }^{\prime }}{%
\partial \phi ^{\prime }}+qr_{2}^{\prime \nu }F_{\nu \mu }^{\prime }=S_{\mu
}^{\sim },  \label{pde}
\end{equation}%
where the source term on the rhs is found to be%
\begin{eqnarray}
S_{\mu }^{\sim } &\equiv &-\left[ \left( u_{0}^{\prime }a^{\prime \alpha
}+u_{\parallel }^{\prime }b^{\prime \alpha }\right) \frac{\partial r_{1\mu
}^{\prime }}{\partial r^{\prime \alpha }}+q\frac{1}{2}r_{1}^{\prime \nu
}r_{1}^{\prime \varsigma }\frac{\partial F_{\varsigma \mu }^{\prime }}{%
\partial r^{\prime \nu }}\right] ^{\sim }  \notag \\
&&-\left[ \frac{\partial r_{1}^{\prime \beta }}{\partial r^{\prime \mu }}%
\left( u_{\beta }^{\prime }-\frac{1}{2}qF_{\beta \nu }^{\prime
}r_{1}^{\prime \nu }\right) \right] ^{\sim }.  \label{s-tilda}
\end{eqnarray}

Notice that the knowledge of $r_{2}^{\prime \nu }$, in analogy with the
corresponding non-relativistic treatment given in Ref.\cite{Cr2013}, permits
one to determine uniquely the second-order perturbation to the GK Lagrangian
form. A preliminary issue concerns the solubility condition of Eq.(\ref{pde}%
), namely the convergence of the solution in both strong and weak electric
field regimes. This can be established by invoking once again the EM-tetrad
representation for $F_{\nu \mu }^{\prime }$ given by Eq.(\ref{tetrad-fmunu})
evaluated at the guiding-center position. It is immediate to prove that the
lhs of Eq.(\ref{pde}) is always defined both when $E\sim 1/O\left(
\varepsilon \right) $ or in the limit $E\rightarrow 0$, so that an exact
solution for $r_{2}^{\prime \nu }$ exists everywhere in the configuration
space for both electric-field regimes.

The solution of Eq.(\ref{pde}) is found to be of the general form%
\begin{equation}
r_{2}^{\prime \mu }=w^{\prime }\left( u_{0}^{\prime }K_{1}^{\prime \mu
}+u_{\parallel }^{\prime }K_{2}^{\prime \mu }\right) +w^{\prime
2}K_{3}^{\prime \mu },  \label{r2-solution-formal}
\end{equation}%
where the coefficients $w^{\prime }$, $u_{0}^{\prime }\ $and $u_{\parallel
}^{\prime }$ are by construction 4-scalars, while $K_{1}^{\prime \mu }$, $%
K_{2}^{\prime \mu }$ and $K_{3}^{\prime \mu }$ are suitable
gyrophase-dependent purely-oscillatory 4-vectors which depend only on the GK
variables $\left( r^{\prime },\phi ^{\prime }\right) $. For the purposes of
the present work the representation (\ref{r2-solution-formal}) is shown to
permit the explicit construction of the GK Lagrangian differential form, and
hence also of the magnetic moment. In fact, this equation displays the
velocity dependences carried by $r_{2}^{\prime \mu }$.

Based on the exact integral representation of the magnetic moment provided
here (see Eq.(\ref{m-exact})), one can show that, by carrying out the
guiding--center perturbative transformation (\ref{1-bis}) and (\ref{2-bis}),
the first order correction $\mu _{1}^{\prime }$ to the magnetic moment is
given by%
\begin{eqnarray}
\mu _{1}^{\prime } &=&\left\langle \frac{\partial r_{2}^{\prime \mu }}{%
\partial \phi ^{\prime }}u_{\mu }^{\prime }\right\rangle _{\phi ^{\prime
}}+\left\langle \frac{\partial r_{1}^{\prime \mu }}{\partial \phi ^{\prime }}%
V_{\mu }^{\prime }\right\rangle _{\phi ^{\prime }}  \notag \\
&&+\frac{1}{2}qF_{\nu \mu }^{\prime }\left\langle r_{1}^{\prime \nu }\frac{%
\partial r_{2}^{\prime \mu }}{\partial \phi ^{\prime }}+r_{2}^{\prime \nu }%
\frac{\partial r_{1}^{\prime \mu }}{\partial \phi ^{\prime }}\right\rangle
_{\phi ^{\prime }}.  \label{mu-1}
\end{eqnarray}%
One can prove that the same result follows equivalently also by direct
calculation in terms of the perturbative Lagrangian differential form,
namely from Eq.(\ref{m-perturb}). Invoking Eqs.(\ref{v-grande-bis}) and (\ref%
{r2-solution-formal}) respectively for $V_{\mu }^{\prime }$ and $%
r_{2}^{\prime \mu }$, one obtains%
\begin{equation}
\mu _{1}^{\prime }=\mu ^{\prime }\left( u_{0}^{\prime }\Delta
_{u_{0}^{\prime }}^{\prime }\left( r^{\prime }\right) +u_{\parallel
}^{\prime }\Delta _{u_{\parallel }^{\prime }}^{\prime }\left( r^{\prime
}\right) \right) +\mu ^{\prime }w^{\prime }\Delta _{w^{\prime }}^{\prime
}\left( r^{\prime }\right) ,  \label{mu1-final-dependence}
\end{equation}%
where the 4-scalar coefficients $\Delta _{u_{0}^{\prime }}^{\prime }\left(
r^{\prime }\right) $, $\Delta _{u_{\parallel }^{\prime }}^{\prime }\left(
r^{\prime }\right) $ and $\Delta _{w^{\prime }}^{\prime }\left( r^{\prime
}\right) $ are only position-dependent. Their precise expression can be
determined after lengthy calculation starting from Eq.(\ref{mu-1}) above.

The following comments are sufficient in order to establish the relevant
qualitative properties of the magnetic moment at this order:

1) The contribution $\mu _{1}^{\prime }$ is linearly proportional to the
leading-order magnetic moment $\mu ^{\prime }$, which is consistent with the
findings of Refs.\cite{Cr2013,Bek1,Bek}.

2)\ Provided $\Delta _{u_{0}^{\prime }}^{\prime }\left( r^{\prime }\right) $
and $\Delta _{u_{\parallel }^{\prime }}^{\prime }\left( r^{\prime }\right) $
are non-zero, the first-order magnetic moment $\mu _{1}^{\prime }$ contains
linear velocity dependences in terms of $u_{0}^{\prime }$ and $u_{\parallel
}^{\prime }$, with $u_{0}^{\prime }$ being given by Eq.(\ref{u00}). This
arises only when first-order perturbative corrections for the 4-velocity are
retained in the guiding-center transformation (\ref{2}) and represents the
novel result of the treatment.

3) In particular, the contribution proportional to $u_{0}^{\prime }$\ is an
intrinsically-relativistic effect since $u_{0}^{\prime }$\ is related to the
other components of the 4-velocity by means of a square-root dependence (see
Eq.(\ref{u00})). In fact, $u_{0}^{\prime }\cong 1$ in the non-relativistic
limit. Concerning the dependences in terms of $u_{\parallel }^{\prime }$, we
notice that besides the linear one, which is analogous to the
non-relativistic result obtained in Ref.\cite{Cr2013}, there is an
additional intrinsically relativistic one appearing through $u_{0}^{\prime }$%
.

4)\ In view of the discussion given above, the representation (\ref%
{mu1-final-dependence}) holds in validity of the strongly-magnetized
ordering irrespective of whether the electric field sub-orderings a) or b)
are invoked.

5)\ The asymptotic approximation $m^{\prime }\cong \mu ^{\prime
}+\varepsilon \mu _{1}^{\prime }$ is an adiabatic invariant of $O\left(
\varepsilon ^{3}\right) $, in the sense that%
\begin{equation}
\frac{1}{\Omega ^{\prime }}\frac{d}{ds}\ln \left[ \mu ^{\prime }+\varepsilon
\mu _{1}^{\prime }\right] =0+O\left( \varepsilon ^{a}\right) ,
\label{ad-inv}
\end{equation}%
with $a\geq 3$.

\section{Relativistic kinetic equilibria}

Based on the results obtained in the previous sections, we are now in
position to start investigating the problem set in the Introduction, namely
the construction of relativistic kinetic equilibria for collisionless
plasmas in curved space-time satisfying the symmetry conditions indicated
above. Following the approach developed in Refs.\cite%
{Cr2010,Cr2012,Cr2011,Cr2011a} to reach the target, the method of invariants
is implemented, which consists in expressing the species KDF in terms of
exact or adiabatic single-particle invariants. In the present case the
latter is identified with the set $(P_{0},m^{\prime })$. Therefore one can
always represent the equilibrium KDF in the form $f=f_{\ast }$, with%
\begin{equation}
f_{\ast }=f_{\ast }\left( \left( P_{0},m^{\prime }\right) ,\Lambda _{\ast
}\right)  \label{f-star}
\end{equation}%
being a smooth strictly-positive function of the particle invariants only
which is summable in velocity-space. Concerning the notation, in Eq.(\ref%
{f-star}) $\left( P_{0},m^{\prime }\right) $ denote explicit functional
dependences, while $\Lambda _{\ast }$ denotes the so-called structure
functions \cite{Cr2011}, namely functions suitably related to the observable
velocity moments of the KDF. These quantities are either constant, i.e. $%
\Lambda _{\ast }=const.$, or more generally functions of the type%
\begin{equation}
\Lambda _{\ast }=\Lambda _{\ast }\left( P_{0},m^{\prime },\varepsilon
^{b}r^{\mu }\right) ,  \label{str-funcdep}
\end{equation}%
where $\varepsilon ^{b}r^{\mu }$ denotes a possible slow dependence and $%
b\geq 1$ is a suitable real number.

In the framework of the perturbative approach developed above, a KDF of the
form (\ref{f-star}) satisfies the Eulerian Vlasov equation%
\begin{equation}
u^{\mu }\frac{\partial f\left( \mathbf{x}\right) }{\partial r^{\mu }}+\frac{%
F^{\mu }}{M_{0}}\frac{\partial f\left( \mathbf{x}\right) }{\partial u^{\mu }}%
=0\left[ 1+O\left( \varepsilon ^{d-1}\right) \right] ,  \label{vlas}
\end{equation}%
where we have denoted $f\left( \mathbf{x}\right) =$ $f_{\ast }$, with $%
f_{\ast }$ prescribed by Eq.(\ref{f-star}), while $F^{\mu }$ is the Lorentz
force due to the external EM field and $d$ is the minimum among the
exponents $a$, $b$ and $k$, where $k$ is the order of adiabatic invariance
for $P_{0}$. Since $m^{\prime }$ is a 4-scalar conserved (in asymptotic
sense) in the EM-tetrad frame, it follows that it is conserved in the same
sense also in the laboratory frame. Therefore, by construction the KDF $%
f_{\ast }$ does not depend explicitly on time in the laboratory frame, at
least in the asymptotic sense indicated also by Eq.(\ref{vlas}). The
previous equation can be equivalently represented in terms of the Lagrangian
equation:%
\begin{equation}
\frac{d}{ds}f\left( \mathbf{x}\left( s\right) \right) =0\left[ 1+O\left(
\varepsilon ^{d-1}\right) \right] ,
\end{equation}%
where $\mathbf{x}\left( s\right) $ is the single-particle phase-space
trajectory. As a consequence, $f\left( \mathbf{x}\left( s\right) \right) $
is an adiabatic invariant of $O\left( \varepsilon ^{d}\right) $ (see also
Eq.(\ref{ad-inv})). Notice that, if for example the exponent $d$ is
determined by the magnetic moment $m^{\prime }$, so that $d\geq 3$, $f_{\ast
}$ satisfies the Vlasov equation (i.e., is conserved) up to proper-time
scales of the order $\tau _{L}/O\left( \varepsilon ^{2}\right) $, where $%
\tau _{L}\equiv 1/\Omega $, and $\Omega $ denoting the Larmor effective
frequency. Notice here that both $\tau _{L}$\ and $\Omega $\ are
dimensionless 4-scalars. Therefore their definition is frame independent. In
practical situations, this means that if for example the invariant parameter 
$\varepsilon $ defined by Eq.(\ref{epsi}) is in the range $10^{-2}-10^{-5}$,
the KDF is conserved up to proper time-scales $10^{4}-10^{10}$ larger than
the Larmor proper-time scale $\tau _{L}$.

A further consequence is that also the velocity moment equations of Eq.(\ref%
{vlas}) are identically satisfied with the same accuracy. It must be
stressed that an equilibrium KDF of the type (\ref{f-star}) represents the
covariant generalization of the corresponding non-relativistic solution
determined in Ref.\cite{Cr2013}. Thus, in principle by proper identification
of the KDF $f_{\ast }$, a perturbative theory analogous to that given in
Refs.\cite{Cr2010,Cr2012,Cr2011,Cr2011a} can be implemented.

For the sake of illustration we consider here a specific possible
realization in which the structure functions are assumed to be constant
functions. An example analogous to those given in Refs.\cite%
{Cr2010,Cr2011,Cr2011a,Cr2012,Cr2013} can be considered, in which $f_{\ast }$
is identified with a generalized Gaussian distribution. To this aim, we
denote with $P_{\mu }=(u_{\mu }+q\frac{A_{\mu }}{\varepsilon })$ the
particle generalized 4-momentum in the observer (laboratory)\ frame which is
characterized by the co-moving 4-velocity $U^{\mu }$, so that in this frame
one has simply $U^{\mu }=\left( 1,0,0,0\right) $. As a consequence, in the
observer frame $P_{\mu }U^{\mu }=P_{0}$ (rest energy), which is a conserved
4-scalar by assumption. Therefore, when evaluated in the EM-tetrad reference
frame the previous scalar can be written as:%
\begin{equation}
P_{0}=P_{\mu R}U_{R}^{\mu }\left( r\right) ,  \label{sollievo}
\end{equation}%
where hereon $P_{\mu R}$ will identify the canonical momentum expressed in
the EM-tetrad reference frame in which the representation%
\begin{equation}
u^{\mu }\equiv u_{0}a^{\mu }+u_{\parallel }b^{\mu }+w\left[ c^{\mu }\cos
\phi +d^{\mu }\sin \phi \right]  \label{vel-eff}
\end{equation}%
holds by assumption. Similarly, $U_{R}^{\mu }\left( r\right) $\ denotes the
4-vector obtained by transforming the co-moving 4-velocity $U^{\mu }$\ to
the same EM-tetrad frame. By construction and in difference with $U^{\mu }$,
one has that $U_{R}^{\mu }\left( r\right) $\ in general exhibits
non-vanishing time- and space-components.

Given these premises, it follows that $f_{\ast }$ can be identified with the
4-scalar%
\begin{equation}
f_{M\ast }=\beta _{\ast }e^{-P_{\mu }U^{\mu }\gamma _{\ast }-m^{\prime
}\alpha _{\ast }},  \label{sol1}
\end{equation}%
where the structure functions are represented by the set of 4-scalars fields 
$\left\{ \Lambda _{\ast }\right\} \equiv \left\{ \beta _{\ast },\gamma
_{\ast },\alpha _{\ast }\right\} $, namely as such they are
frame-independent. As a general feature of the present solution method, it
is important to remark that the freedom in the prescription of the structure
functions, which are in principle of the general type (\ref{str-funcdep}),
makes possible the practical application of the theory to model concrete
physical configurations or to fit real observational data. In particular,
from the physical point of view, the meaning of the structure functions $%
\Lambda _{\ast }$ is that they are associated with the relevant fluid fields
of the system, and therefore with physical observables (see also related
discussion in Refs.\cite{Cr2011,Cr2011a,Cr2013} for the corresponding
non-relativistic interpretation). In detail, here $\beta _{\ast }$\ is
related to the plasma 4-flow, or equivalently the plasma number density when
measured in the fluid co-moving frame, while $\gamma _{\ast }$\ and $\alpha
_{\ast }$\ are related to the temperature anisotropy, namely to the parallel
and perpendicular temperatures of the plasma defined with respect to the
local direction of magnetic field. For an illustration of the issue and in
agreement with the scope of the study, for simplicity $\Lambda _{\ast }$\
will be considered constant in the following.

We stress that the representation of the KDF in Eq.(\ref{sol1}) is still
exact, in the sense that no asymptotic approximations have been introduced
there. Therefore, also the magnetic moment $m^{\prime }$ in the exponential
factor must be regarded as given by its exact, i.e., non-perturbative,
representation by Eq.(\ref{m-exact}). On the other hand, in view of these
assumptions, after invoking Eq.(\ref{sollievo}) to express the 4-scalar $%
P_{\mu }U^{\mu }$ in the EM-tetrad frame and the perturbative representation
(\ref{m-perturb}) for $m^{\prime }$\ obtained in the framework of the
perturbative GK theory, the equilibrium KDF correct to first-order in $%
\varepsilon $ takes the form%
\begin{equation}
f_{M\ast }=\beta _{\ast }e^{-P_{\mu R}U_{R}^{\mu }\left( r\right) \gamma
_{\ast }-\mu ^{\prime }\alpha _{\ast }}\left[ 1-\varepsilon \mu _{1}^{\prime
}\alpha _{\ast }\right] .  \label{sol-2}
\end{equation}%
We remark that this KDF is summable. To warrant that it is also strictly
positive we notice that the perturbation $\varepsilon \mu _{1}^{\prime
}\alpha _{\ast }$ can only be treated as being an infinitesimal. In a strict
sense this happens only in the subset of phase-space in which $\varepsilon
\mu _{1}^{\prime }\alpha _{\ast }\sim O\left( \varepsilon \right) $. Here we
notice that the two terms in the exponential are still evaluated
respectively at the actual particle position and the corresponding
guiding-center position (both defined with respect to the EM-tetrad
reference frame).

In principle, the previous form of the KDF lends itself to determine
explicitly 4-velocity moments. However, for this purpose the magnetic moment 
$m^{\prime }$ must be preliminarily evaluated at the actual particle
position. This requires the introduction of the inverse guiding-center
transformation for the GK state $\mathbf{z}^{\prime }\equiv \left( \mathbf{y}%
^{\prime },\phi ^{\prime }\right) $ of the type $\mathbf{z}^{\prime }=%
\mathbf{z}^{\prime }\left( \mathbf{x}\right) $ \cite{Cr2013}. When this is
applied to the perturbative representation of $m^{\prime }$ given by Eq.(\ref%
{m-perturb}), such a back-transformation can be carried out explicitly in
principle to arbitrary order in $\varepsilon $. In particular, correct
through $O\left( \varepsilon \right) $, this leads to the following
expression for the magnetic moment $m=m^{\prime }\left( \mathbf{x}\right) $
evaluated at the actual particle position:%
\begin{equation}
m=\mu +\varepsilon \mu _{1}+\varepsilon \delta _{\left( \mu \right) }.
\label{mmm}
\end{equation}%
Here $\mu \equiv \frac{w^{2}}{2qH}$ is the leading-order contribution,
while, from Eq.(\ref{mu1-final-dependence}), the first-order term $\mu _{1}$
is simply%
\begin{equation}
\mu _{1}=\mu \left( u_{0}\Delta _{u_{0}}\left( r\right) +u_{\parallel
}\Delta _{u_{\parallel }}\left( r\right) \right) +\mu w\Delta _{w}\left(
r\right) .
\end{equation}%
Finally, the $O\left( \varepsilon \right) $ correction $\delta _{\left( \mu
\right) }$ originates from the inverse guiding-center transformation applied
to $\mu ^{\prime }$ and this term is comparable in order of magnitude with $%
\varepsilon \mu _{1}$. It is sufficient to notice here that, from the
second-order Larmor-radius expansion, the functional dependence of $\delta
_{\left( \mu \right) }$ must be generally of the type $\delta _{\left( \mu
\right) }=\delta _{\left( \mu \right) }\left( r,u_{\parallel },\mu ,\phi
\right) $, thus including also explicit gyrophase dependences.

In terms of Eq.(\ref{mmm}) and neglecting second-order corrections in $%
\varepsilon $, the equilibrium KDF (\ref{sol-2}) becomes%
\begin{equation}
f_{M\ast }=\beta _{\ast }e^{-P_{\mu R}U_{R}^{\mu }\left( r\right) \gamma
_{\ast }-\mu \alpha _{\ast }}\left[ 1-\left( \varepsilon \mu
_{1}+\varepsilon \delta _{\left( \mu \right) }\right) \alpha _{\ast }\right]
,  \label{result}
\end{equation}%
where all quantities are represented in the EM-tetrad with origin at the
actual particle position. In view of the theory developed above, Eq.(\ref%
{result}) is a possible realization of Vlasov kinetic equilibria in the
presence of spatially non-symmetric systems characterized by stationary
gravitational and EM fields when referred to the observer reference frame.
Notice that in Eq.(\ref{result}) the kinematic constraint $u^{\mu }u_{\mu
}=1 $\ has not been imposed yet.

To conclude this analysis, it is important to remark that, despite the
formal simplicity of the solution (\ref{sol1}) and its asymptotic
approximations (\ref{sol-2}) and (\ref{result}), the theory developed here
is intrinsically sophisticated in several respects. The complexity of the
formalism arises as a distinct feature of non-isotropic collisionless
plasmas in comparison with neutral gases or isotropic Maxwellian-like
plasmas. It is immediate to realize that this feature is carried in
particular by the magnetic moment and is \textquotedblleft
hidden\textquotedblright\ in its appropriate determination in the framework
of the covariant GK theory developed above, in turn based on the EM-tetrad
formalism in curved space-time.

\subsection{The fluid 4-flow}

The functional dependences contained in the KDF suggest us that it can
sustain fluid equilibria characterized by non-uniform 4-flows and
non-isotropic momentum-energy tensor. In particular, the 4-flow $N^{\mu
}\left( r\right) $ is defined in terms of $f_{\ast }$ as the 4-velocity
integral%
\begin{equation}
N^{\mu }\left( r\right) =2c\int \sqrt{-g}d^{4}u\Theta \left( u^{0}\right)
\delta \left( u^{\mu }u_{\mu }-1\right) u^{\mu }f_{M\ast },
\label{fluid-4-flow-uno}
\end{equation}%
where as usual the Dirac-delta takes into account the kinematic constraint
for the 4-velocity. Invoking Eq.(\ref{fluid-4-flow-uno}) the integral becomes%
\begin{equation}
N^{\mu }\left( r\right) =c\int \frac{\sqrt{-g}d^{3}u}{\sqrt{1+u_{\parallel
}^{2}+w^{2}}}u^{\mu }f_{M\ast }.  \label{enne-mu}
\end{equation}%
Now we notice that the previous integral is evaluated with respect to the
EM-tetrad reference frame, in which locally, thanks to the principle of
equivalence, $\sqrt{-g}=1$. Hence, let us introduce the cylindrical
coordinates in the velocity space:%
\begin{equation}
\int d^{3}u\rightarrow \int_{0}^{2\pi }d\phi \int_{0}^{+\infty
}wdw\int_{-\infty }^{+\infty }du_{\parallel },  \label{volume-fluid}
\end{equation}%
where $u_{\parallel }$ and $w$ coincide with the scalar components of the
4-velocity given by Eq.(\ref{vel-eff}), in terms of which the KDF is
represented (see Eq.(\ref{result})). Hence, the integral becomes%
\begin{equation}
N^{\mu }\left( r\right) =c\int_{0}^{2\pi }d\phi \int_{0}^{+\infty
}wdw\int_{-\infty }^{+\infty }du_{\parallel }\frac{u^{\mu }f_{M\ast }}{\sqrt{%
1+u_{\parallel }^{2}+w^{2}}},
\end{equation}%
whose explicit (numerical) evaluation is in principle straightforward.

Once $u^{\mu }$ is represented in terms of Eq.(\ref{vel-eff}), the general
form of $N^{\mu }\left( r\right) $ becomes therefore of the type%
\begin{equation}
N^{\mu }\left( r\right) =N_{0}a^{\mu }+N_{\parallel }b^{\mu }+N_{\perp
1}c^{\mu }+N_{\perp 2}d^{\mu },  \label{www}
\end{equation}%
where the coefficients $N_{0}$, $N_{\parallel }$, $N_{\perp 1}$ and $%
N_{\perp 2}$ are 4-scalars which are generally position-dependent. Hence
this equation permits to determine explicitly $N^{\mu }\left( r\right) $ in
arbitrary reference frames (coordinate-systems). In particular, we notice
that, thanks to the time-symmetry property of the equilibrium KDF in the
laboratory frame, also $N^{\mu }\left( r\right) $ cannot depend explicitly
on the time coordinate when represented in the same frame.

The result permits us also to cast some light on the qualitative\ physical
properties exhibited by $N^{\mu }\left( r\right) $.

First, there are only three physical mechanisms responsible for the
generation of the 4-flow, the first two being of $O\left( \varepsilon
^{0}\right) $:

1) the first one is produced by the boost 4-velocity $U_{R}^{\mu }\left(
r\right) $. This effect can become dominant in the case of intense
gravitational and EM fields, in which $U_{R}^{\mu }\left( r\right) $ is in
fact relativistic \cite{Bek};

2)\ the second one is associated with the velocity-space anisotropy due to $%
\mu $, which is expected to generate temperature anisotropy. For comparison
see Refs.\cite{Cr2011a,Cr2013}, where this effect is considered in the
non-relativistic context;

3)\ the last one is provided by the first-order Larmor-radius corrections
due to $\varepsilon \mu _{1}$ and $\varepsilon \delta _{\left( \mu \right) }$%
.

Notice that the first-order perturbations cannot generally be neglected.
This happens if some of the leading-order components (of $N^{\mu }\left(
r\right) $) vanishes locally. A situation of this type occurs for example in
the non-relativistic treatment of these equilibria, as pointed out in Ref.%
\cite{Cr2013}. We also notice that in case the structure functions are taken
to be non-constant, then in principle additional first-order contributions
may be expected to arise in the fluid 4-flow.

Second, we stress that the resulting 4-flow in all cases satisfies by
construction the fluid conservation law, which within the present
perturbative theory becomes $\nabla _{\mu }N^{\mu }\left( r\right)
=0+O\left( \varepsilon ^{c}\right) $.

Third, let us consider the spatial dependences in terms of $r^{\mu }$
arising in $N^{\mu }\left( r\right) $. In the case in which the structure
functions are constant (as it is the case here), non-trivial
configuration-space dependences still arise due to the following physical
effects:\ 1) the explicit dependence in terms of the 4-scalar $A_{\mu
}U_{R}^{\mu }\left( r\right) $ associated with $P_{\mu }$; 2)\ the
functional form of the 4-vector $U_{R}^{\mu }\left( r\right) $, which is
determined by the boost transformation; and finally 3) the spatial
dependences appearing in the 4-scalars $\mu $, $\mu _{1}$ and $\delta
_{\left( \mu \right) }$ occurring due to the inhomogeneities of the
background EM field.

Finally, we must stress that the present kinetic theory can in principle be
generalized to the case of non-constant structure functions. Such a
development can be achieved along the lines pointed out in Refs.\cite%
{Cr2010,Cr2012,Cr2011,Cr2011a}.

\subsection{The non-relativistic limit}

It is interesting to consider behavior of the solution obtained in this
section in the non-relativistic limit. This limit is realized when the
system exhibits both non--relativistic flow velocities and temperatures and
the gravitational field can be treated within the classical Newtonian
theory. In such a case the non-relativistic Vlasov kinetic equation $\frac{d%
}{dt}f\left( \mathbf{x}\left( t\right) \right) =0$ can be used as the
dynamic equation for the KDF, where now $t$ denotes the absolute time, to be
identified with the observer time coordinate. Similarly, single-particle
dynamics can be described by a non-relativistic Lagrangian function, in
terms of which the corresponding non-relativistic GK theory can be
consistently formulated (see Refs.\cite{Cr2011,Cr2011a,Cr2013}).

In this framework, for a particle characterized by mass $M$ and charge $Ze$
(species index is omitted for simplicity), the stationarity condition
implies the conservation of the total particle energy $E$, namely%
\begin{equation}
E=\frac{Mv^{2}}{2}+Ze\Phi ^{eff}\equiv {Ze}\Phi _{\ast },  \label{energii}
\end{equation}%
where $\Phi ^{eff}\equiv \Phi +\frac{M}{Ze}\Phi _{G}$, with $\Phi $ and $%
\Phi _{G}$ being the electrostatic (ES) and gravitational potential
respectively. In this regard we notice that such a type of description can
be extended in principle to include also the case of strong gravity regimes,
where the space-time can be described by means of pseudo-Newtonian
potentials \cite{PW,Z08,dt,mano}. In the absence of global spatial
symmetries the only additional adiabatic invariant is represented by the
magnetic moment. Its perturbative representation is still given formally by
Eq.(\ref{m-perturb}) above, while the explicit calculation of the series
coefficients in the non-relativistic case is reported in Ref.\cite{Cr2013}.

As a result, the non-relativistic representation of the equilibrium solution
(\ref{sol1}) is given by the following KDF%
\begin{equation}
f_{M\ast }=\beta _{\ast }e^{-{Ze}\Phi _{\ast }\gamma _{\ast }-m^{\prime
}\alpha _{\ast }},  \label{sol1-nr}
\end{equation}%
where again the set of structure functions are identified with $\left\{
\Lambda _{\ast }\right\} \equiv \left\{ \beta _{\ast },\gamma _{\ast
},\alpha _{\ast }\right\} $ and can be taken to be identically constant. In
particular, following Ref.\cite{Cr2013}, it is immediate to prove that $%
\beta _{\ast }$ is related to the non-uniform number density, $\gamma _{\ast
}\equiv \frac{1}{T_{\parallel }}$ represents the parallel plasma temperature
(defined in a statistical way with respect to the local direction of the
magnetic field), while $\alpha _{\ast }\equiv \frac{B}{\Delta _{T}}$ carries
the information about the temperature anisotropy $\frac{1}{\Delta _{T}}%
\equiv \frac{1}{T_{\perp }}-\frac{1}{T_{\parallel }}$, where $T_{\perp }$ is
the perpendicular temperature and $B$ the magnitude of the magnetic field.

Besides the intrinsic different treatment of the gravitational field with
respect to the covariant formalism, other relevant departures between the
relativistic and non-relativistic kinetic equilibria are:

1) The solution (\ref{sol1-nr}) is defined in the phase-space spanned by the
position and velocity 3-vectors, while Eq.(\ref{sol1}) is defined in the
extended phase-space spanned by the position and velocity 4-vectors, with
the 4-velocity being subject to the mass-shell condition (\ref{mass-shell}).

2)\ In the non-relativistic limit, the EM-tetrad formalism is replaced by
the introduction of an absolute magnetic-related coordinate system defined
by the parallel and perpendicular directions to the local magnetic field
line.

3) From the same definition of EM-tetrad frame, the boost 4-velocity $%
U_{R}^{\mu }\left( r\right) $ is uniquely defined in the non-relativistic
framework and reduces to the so-called $\left( \mathbf{E}\times \mathbf{B}%
\right) $-drift velocity, to be properly generalized with the inclusion of
the contribution arising from the gravitational potential \cite%
{Cr2011,Cr2013,bek3}.

As a final discussion, let us consider the change occurring in the
calculation of the 4-flow for the non-relativistic solution. In such a case
the time and space components of $N^{\mu }\left( r\right) $ identify
uniquely the number density $n$ and the fluid flow velocity $\mathbf{V}$ of
the system, so that%
\begin{equation}
N^{\mu }\left( r\right) \rightarrow \left( n,\mathbf{V}\right) .
\end{equation}%
Spatial dependences in the leading-order number density arise in this limit
due the effective potential $\Phi ^{eff}$ and the $\left( \mathbf{E}\times 
\mathbf{B}\right) $-drift velocity. More interesting is the analysis of the
properties of the flow velocity. As shown in Ref.\cite{Cr2013}, in the
magnetic-related coordinate system, the relevant contributions become%
\begin{equation}
\mathbf{V}=\mathbf{V}_{\perp }+\varepsilon V_{\parallel }\mathbf{b},
\end{equation}%
where $\mathbf{b}\equiv \mathbf{B}/|\mathbf{B}|$ is the unit vector parallel
to the magnetic field and $\varepsilon $ denotes the non-relativistic
Larmor-radius expansion parameter (see again Ref.\cite{Cr2013} for its
definition). Here, the only leading-order contribution is $\mathbf{V}_{\perp
}$, which is perpendicular to the magnetic field direction and is found to
be proportional to the same drift velocity, while it vanishes for isotropic
temperature. This component is the non-relativistic analogue of the
contribution associated with the boost 4-velocity $U_{R}^{\mu }\left(
r\right) $ in the case of the 4-flow $N^{\mu }\left( r\right) $. However, in
the non-relativistic limit and in difference with the conclusions pointed
out after Eq.(\ref{www}) above, the only parallel component of the flow
arises as a first-order correction, denoted as $V_{\parallel }$. This is
found to be exclusively generated by the corresponding first-order
contributions associated with the adiabatic conservation of the magnetic
moment, which therefore appears to be the only physical mechanism for the
occurrence of macroscopic parallel flows in non-relativistic and spatially
non-symmetric collisionless plasmas. Therefore, in the non-relativistic
limit the admissible phenomenology related to the possible occurrence of
plasma flows in these systems appears more restricted and constrained with
respect to the relativistic case.

\section{Conclusions}

In this paper two distinct but intimately related topics have been
investigated. The first one concerns the construction of the relativistic
magnetic moment characterizing charged particle dynamics in the presence of
intense gravitational and EM fields which satisfy suitable asymptotic
orderings. The second one deals instead with the construction of kinetic
equilibria for the Vlasov equation governing relativistic plasmas immersed
in configurations characterized by gravitational and EM fields which are
stationary\ with respect to the observer but at the same time do not exhibit
any kind of spatial symmetry.

The gyrokinetic (GK) theory formulated here is developed in the context of a
non perturbative approach and is based on an extended guiding-center
transformation which includes also a corresponding relativistic 4-velocity
transformation. As a consequence, this feature warrants the validity of the
theory even in the case of weak or locally vanishing electric field.

A remarkable feature of the perturbative theory is that it applies under
quite general conditions, namely:\textbf{\ }1) in the absence of any kind of
symmetry requirement on the background fields,\textbf{\ }2) without imposing
restrictions on possible fast space-time dependences of the EM field,
albeit, as long they can be considered slow with respect to Larmor rotation
dynamics and,\textbf{\ }3) fulfilling all the required physical constraints,
i.e., in particular the extremal relationship between 4-velocity and
4-position perturbations (see the constraint equation Eq.(\ref{ext-1})).
Instead, the mass-shell condition (\ref{mass-shell}) is satisfied
automatically by the extremal 4-velocity (i.e., which is determined by means
of the Euler-Lagrange equations following from the Hamilton variational
principle).

In order to implement these features,\textbf{\ }a variational approach to GK
theory based on the relativistic Hamilton variational principle is adopted
which is expressed in terms of super-abundant and hybrid (i.e., generally
non-canonical) gyrokinetic variables. It is shown that in such a case the
gyrokinetic transformation, both finite or infinitesimal, is subject to
physical realizability conditions, which are mandatory for the existence of
the transformation itself. In particular, when the perturbative expansion is
carried out in terms of a Larmor-radius expansion, explicit solution to the
relevant order of the perturbative guiding-center and the GK transformations
have been displayed. A further notable consequence is the asymptotic
convergence of the theory, which has been proved to hold for
strongly-magnetized plasmas in both strong and weak electric field regimes.
As a main result, the theory permits one to determine both exact and
perturbative\ representations for the particle guiding-center magnetic
moment, in terms of the Larmor-radius expansion parameter $\varepsilon $.

Explicit realizations of kinetic equilibria, i.e., stationary solutions of
the relativistic Vlasov equation, have been constructed by implementing the
method of invariants. These have been identified with the canonical momentum
associated with the time symmetry (i.e., the stationarity condition) and the
particle magnetic moment. In particular, the case of Gaussian-like
distributions has been considered. Kinetic equilibria of this type have been
shown to be generally non-uniform in space-time. The knowledge of such
equilibrium kinetic distribution functions allows one to determine
consistently also the corresponding fluid/MHD description. In this study we
have considered as an example the fluid 4-flow, by pointing out its basic
qualitative physical properties. An explicit formal representation for the
fluid 4-flow has been determined in terms of its EM-tetrad representation.
As a result, we have shown that, to leading-order in $\varepsilon $, 4-flows
can be generated in particular both by the boost 4-velocity carried by the
EM-tetrad and the magnetic moment conservation. Additional effects, which
are of first-order in $\varepsilon $ and are uniquely associated with the
magnetic moment, have been pointed out.

The kinetic theory developed here provides the theoretical framework for the
statistical description of astrophysical collisionless plasmas arising in
relativistic regimes and subject to the simultaneous action of intense
gravitational and electromagnetic fields. The remarkable feature pointed out
here is that such equilibria can actually sustain relativistic flows in
spatially non-symmetric scenarios. Such features are expected to be
ubiquitous in a host of astrophysically-relevant situations. Typical
examples are provided by non-symmetric disc-jet structures in the
surrounding of compact objects, like stellar-mass and galactic-center black
holes or active galactic nuclei (AGNs), where high-energy plasmas are likely
to be generated.

\textbf{Acknowledgments - }Financial support by the Italian Foundation
\textquotedblleft Angelo Della Riccia\textquotedblright\ (Firenze, Italy) is
acknowledged by C.C. Work developed within the research projects of the
Czech Science Foundation GA\v{C}R grant No. 14-07753P (C.C.)\ and Albert
Einstein Center for Gravitation and Astrophysics, Czech Science Foundation
No. 14-37086G (M.T. and Z.S.). C.C. would like to thank Dr. Remo Fulvio
Gavazzoni (Consultant Dermatologist at the Clinical Institute
\textquotedblleft Citt\`{a} di Brescia\textquotedblright , Brescia, Italy)
for his help during the preparation of this manuscript.


\begin{thebibliography}{99}
\bibitem{LL} L.D. Landau and E.M. Lifschitz, \textit{Field theory,
Theoretical Physics Vol.2} (Addison-Wesley, N.Y., 1957).

\bibitem{Cr2010} C. Cremaschini, J.C. Miller and M. Tessarotto, Phys.
Plasmas \textbf{17}, 072902 (2010).

\bibitem{Cr2012} C. Cremaschini and M. Tessarotto, Phys. Plasmas \textbf{19}%
, 082905 (2012).

\bibitem{maha-1} V.I. Berezhiani, R.D. Hazeltine and S.M. Mahajan, Phys.
Rev. E \textbf{69}, 056406 (2004).

\bibitem{maha-2} R.D. Hazeltine and S.M. Mahajan, Phys. Rev. E \textbf{70},
036404 (2004).

\bibitem{EPJ1} C. Cremaschini and M. Tessarotto, Eur. Phys. J. Plus \textbf{%
126}, 42 (2011).

\bibitem{EPJ2} C. Cremaschini and M. Tessarotto, Eur. Phys. J. Plus \textbf{%
126}, 63 (2011).

\bibitem{EPJ3} C. Cremaschini and M. Tessarotto, Eur. Phys. J. Plus \textbf{%
127}, 4 (2012).

\bibitem{EPJ4} C. Cremaschini and M. Tessarotto, Eur. Phys. J. Plus \textbf{%
127}, 103\textbf{\ }(2012).

\bibitem{EPJ5} C. Cremaschini and M. Tessarotto, Phys. Rev. E \textbf{87},
032107 (2013).

\bibitem{EPJ6} C. Cremaschini and M. Tessarotto, Int. J. Mod. Phys. A 
\textbf{28}, 1350086 (2013).

\bibitem{cz1} A. Janiuk, B. Czerny, Mon. Not. R. Ast. Soc. \textbf{414},
2186 (2011).

\bibitem{as7} J. Hor\'{a}k and V. Karas, Mon. Not. R. Ast. Soc. \textbf{365}%
, 813 (2006).

\bibitem{as8} V. Karas, Astronomische Nachrichten \textbf{327}, 961 (2006).

\bibitem{as2} D. Bini, R.T. Jantzen and L. Stella, Class. Quantum Grav. 
\textbf{26}, 055009 (2009).

\bibitem{as3} D. Bini, A. Geralico, R.T. Jantzen, O. Semer\'{a}k and L.
Stella, Class. Quantum Grav. \textbf{28}, 035008 (2011).

\bibitem{as1} D. Bini, M. Falanga, A. Geralico and L. Stella, Class. Quantum
Grav. \textbf{29}, 065014 (2012).

\bibitem{as9} M. Dov\v{c}iak, F. Muleri, R.W. Goosmann, V. Karas and G.
Matt, Mon. Not. R. Ast. Soc. \textbf{391}, 32 (2008).

\bibitem{as10} M. Dov\v{c}iak, F. Muleri, R.W. Goosmann, V. Karas and G.
Matt, Astrophys. J. \textbf{731}, 75 (2011).

\bibitem{ks1} M. Kolo\v{s} and Z. Stuchl\'{\i}k, Phys. Rev. D \textbf{82},
125012 (2010).

\bibitem{ks2} Z. Stuchl\'{\i}k and M. Kolo\v{s}, Phys. Rev. D \textbf{85},
065022 (2012).

\bibitem{ks3} Z. Stuchl\'{\i}k and M. Kolo\v{s}, JCAP \textbf{10}, 08 (2012).

\bibitem{ks4} M. Kolo\v{s} and Z. Stuchl\'{\i}k, Phys. Rev. D \textbf{88},
065004 (2013).

\bibitem{ks5} A. Tursunov, M. Kolo\v{s}, B. Ahmedov, Z. Stuchl\'{\i}k, Phy.
Rev. D \textbf{87}, 125003 (2013).

\bibitem{ks6} J. Kov\'{a}\v{r}, Eur. Phys. J. Plus \textbf{128}, 142 (2013).

\bibitem{ks7} Z. Stuchl\'{\i}k and M. Kolo\v{s}, Phys. Rev. D \textbf{89},
065007 (2014).

\bibitem{as6} M. Falanga, L. Kuiper, J. Poutanen, D.K. Galloway, E.W.
Bonning, E. Bozzo, A. Goldwurm, W. Hermsen, L. Stella, Astron. Astrophys. 
\textbf{529}, A68 (2011).

\bibitem{as4} M. Falanga, L. Kuiper, J. Poutanen, D.K. Galloway, E. Bozzo,
A. Goldwurm, W. Hermsen and L. Stella, Astron. Astrophys. \textbf{545}, A26
(2012).

\bibitem{co2} F.K. Baganoff, Y. Maeda, M. Morris, M.W. Bautz, W.N. Brandt,
W. Cui, J.P. Doty, E.D. Feigelson, G.P. Garmire, S.H. Pravdo, G.R. Ricker,
L.K. Townsley, Astrophys. J. \textbf{591}, 891 (2003).

\bibitem{co5} G. Bower, M.C.H. Wright, H. Falcke and D. Backer, Astrophys.
J. \textbf{588}, 331 (2003).

\bibitem{co3} D.P. Marrone, J.M. Moran, J.H. Zhao and R. Rao, Astrophys. J. 
\textbf{640}, 308 (2005).

\bibitem{co1} T. Islam, Astrophys. J. \textbf{746}, 8 (2012).

\bibitem{co4} K. Faghei, Astrophysics and Space Science \textbf{345}, 125
(2013).

\bibitem{degroot} S.R. De Groot, \textit{Relativistic Kinetic Theory:
Principles and Applications, }Elsevier Science Ltd (1980), ISBN-13:
978-0444854537.

\bibitem{Cr2011} C. Cremaschini, J.C. Miller and M. Tessarotto, Phys.
Plasmas \textbf{18}, 062901 (2011).

\bibitem{Cr2011a} C. Cremaschini and M. Tessarotto, Phys. Plasmas \textbf{18}%
, 112502 (2011).

\bibitem{PRL} C. Cremaschini, M. Tessarotto and J.C. Miller, Phys. Rev.
Lett. \textbf{108}, 101101 (2012).

\bibitem{Cr2013} C. Cremaschini and M. Tessarotto, Phys. Plasmas \textbf{20}%
, 012901 (2013).

\bibitem{Cr2013b} C. Cremaschini and Z. Stuchl\'{\i}k, Phys. Rev. E \textbf{%
87}, 043113 (2013).

\bibitem{Cr2013c} C. Cremaschini, Z. Stuchl\'{\i}k and M. Tessarotto, Phys.
Plasmas \textbf{20}, 052905 (2013).

\bibitem{PRE-new} C. Cremaschini, Z. Stuchl\'{\i}k and M. Tessarotto, Phys.
Rev. E \textbf{88}, 033105 (2013).

\bibitem{Catto1987} P.J. Catto, I.B. Bernstein and M. Tessarotto, Phys.
Fluids B \textbf{30}, 2784 (1987).

\bibitem{Volpi} D. Volpi, L. Del Zanna, E. Amato and N. Bucciantini, Astron.
Astrophys. \textbf{485}, 337 (2008).

\bibitem{APJS} C. Cremaschini, J. Kov\'{a}\v{r}, P. Slan\'{y}, Z. Stuchl%
\'{\i}k and V. Karas, Astrophys. J. Suppl. \textbf{209}, 15 (2013).

\bibitem{pe-1} G.G. Howes, W. Dorland, S.C. Cowley, G.W. Hammett, E.
Quataert, A.A. Schekochihin and T. Tatsuno, Phys. Rev. Lett. \textbf{100},
065004 (2008).

\bibitem{pe0} L. Sironi and A. Spitkovsky, Astrophys. J. \textbf{698}, 1523
(2009).

\bibitem{pe1} R.F. Penna, J.C. McKinney, R. Narayan, A. Tchekhovskoy, R.
Shafee and J.E. McClintock, Mon. Not. R. Ast. Soc. \textbf{408}, 752 (2010).

\bibitem{pe2} R. Narayan, A. Sadowski, R.F. Penna and A.K. Kulkarni,
Mon. Not. R. Ast. Soc. \textbf{426}, 3241 (2012).

\bibitem{pe3} R.F. Penna, A. Kulkarni and R. Narayan, Astron. Astrophys. 
\textbf{559}, A116 (2013).

\bibitem{pe4} L. Sironi, A. Spitkovsky, J. Arons, Astrophys. J. \textbf{771}%
, 54 (2013).

\bibitem{pe5} E. Tejeda, P.A. Taylor and J.C. Miller, Mon. Not. R. Ast. Soc. 
\textbf{429}, 925 (2013).

\bibitem{pe6} L. Naso, W. Klu\'{z}niak and J.C. Miller, Mon. Not. R. Ast.
Soc. \textbf{435}, 2633 (2013).

\bibitem{ma-1} S.M. Mahajan, Phys. Rev. Lett. \textbf{90}, 035001 (2003).

\bibitem{ma-2} A.R. Soto-Chavez, S.M. Mahajan and R.D. Hazeltine, Phys. Rev.
E \textbf{81}, 026403 (2010).

\bibitem{ma-3} F.A. Asenjo, V. Mu\~{n}oz, J.A. Valdivia and T. Hada, Phys.
Plasmas \textbf{16}, 122108 (2009).

\bibitem{ma-4} J. Pino, H. Li and S. Mahajan, Phys. Plasmas \textbf{17},
112112 (2010).

\bibitem{Re2} K. Dionysopoulou, D. Alic, C. Palenzuela, L. Rezzolla and B.
Giacomazzo, Phys. Rev. D \textbf{88}, 044020 (2013).

\bibitem{Re1} F. Galeazzi, W. Kastaun, L. Rezzolla and J.A. Font, Phys. Rev.
D \textbf{88}, 064009 (2013).

\bibitem{Re0} N. Bucciantini and L. Del Zanna, Mon. Not. R. Ast. Soc. 
\textbf{428}, 71 (2013).

\bibitem{Re00} O. Korobkin, E. Abdikamalov, N. Stergioulas, E. Schnetter, B.
Zink, S. Rosswog, C.D. Ott, Mon. Not. R. Ast. Soc. \textbf{431}, 349 (2013).

\bibitem{na2} M. Machida and R. Matsumoto, Astrophys. J. \textbf{585},
429--442 (2003).

\bibitem{na3} H. Kigure and K. Shibata, Astrophys. J. \textbf{634}, 879-900
(2005).

\bibitem{na4} J.C. McKinney and R.D. Blandford, Mon. Not. R. Ast. Soc. 
\textbf{394}, L126 (2009).

\bibitem{na5} B. Punsly, I.V. Igumenshchev and S. Hirose, Astrophys. J. 
\textbf{704}, 1065-1085 (2009).

\bibitem{na6} J.C. McKinney, A. Tchekhovskoy and R.D. Blandford, Mon. Not.
R. Ast. Soc. \textbf{423}, 3083 (2012).

\bibitem{na1} O. Porth, Mon. Not. R. Ast. Soc. \textbf{429}, 2482 (2013).

\bibitem{Grad84} H. Grad, Proc. of the Workshop on mathematical aspects of
fluid and plasma dynamics, Trieste, Italy, May 30 - June 02, 1984, Ed. C.
Cercignani, S. Rionero and M. Tessarotto (Quaderni del CNR, Gruppo Nazionale
per la Fisica Matematica, 1984).

\bibitem{Grad67} H. Grad, Phys. Fluids \textbf{10}, 137-154 (1967).

\bibitem{Grad58} H. Grad and H. Rubin, Proc. Second United Nation Conference
on the Peaceful Uses of Atomic Energy, Geneva 1958, vol.31, pp.190-197
(1958).

\bibitem{Bek1} A. Beklemishev and M. Tessarotto, Phys. Plasmas \textbf{6},
4487 (1999).

\bibitem{Bek} A. Beklemishev and M. Tessarotto, Astron. Astrophys. \textbf{%
428}, 1 (2004).

\bibitem{bek3} C. Cremaschini, M. Tessarotto and Z. Stuchl\'{\i}k, Phys.
Plasmas \textbf{21}, 032902 (2014).

\bibitem{ehl} J. Ehlers, Lecture Notes in Physics \textbf{28}, 78 (1974).

\bibitem{lit81} R.G. Littlejohn, Phys. Fluids\textbf{\ 24}, 1730 (1981).

\bibitem{hahm} T.S. Hahm, W.W. Lee and A. Brizard, Phys. Fluids \textbf{31},
1940 (1988).

\bibitem{grav} C.W. Misner, K.S. Thorne and J.A. Wheeler, \textit{Gravitation%
} (Freeman, San Francisco, 1973).

\bibitem{PW} B. Paczynsky and P.J. Wiita, Astron. Astrophys. \textbf{88}, 23
(1980).

\bibitem{Z08} Z. Stuchl\'{\i}k and J. Kov\'{a}\v{r}, Int. J. Mod. Phys. D 
\textbf{17}, 2089 (2008).

\bibitem{dt} Z. Stuchl\'{\i}k, P. Slan\'{y}, and J. Kov\'{a}\v{r}, Classical
Quant. Grav. \textbf{26}, 215013 (2009).

\bibitem{mano} E. Tejeda and S. Rosswog, MNRAS \textbf{433}, 1930-1940
(2013).
\end{thebibliography}
\end{document}